\documentclass[a4paper,11pt]{scrartcl}

\usepackage[english]{babel} 

\usepackage{graphicx} 
\usepackage{amsmath,amsthm,amssymb} 
\usepackage[utf8]{inputenc} 

\usepackage[margin=1.8cm,footskip=0.1cm]{geometry}

\usepackage{setspace}
\usepackage{lastpage}
\usepackage{fancyhdr}
\usepackage{pdflscape}

\usepackage[labelfont=bf, labelsep=colon, format=plain]{caption}

\usepackage[square,numbers]{natbib}
\usepackage[section, below]{placeins}

\newcommand{\myvec}[1]{\mathbf{#1}}
\newcommand{\mymat}[1]{\mathbf{#1}}
\newcommand{\myref}[2]{\textsc{#1}\ref{#2}}

\title{Modeling the timing of 3D gradient echo BOLD fMRI data}
\author{}
\date{}

\begin{document}

\thispagestyle{empty}
\begin{center}
    \section*{Effects of Sequence Timing on the Spatio-Temporal Properties of 3D BOLD fMRI: A Formal Framework and Analysis}
\end{center}
\noindent \textbf{Samuel Bianchi}\textsuperscript{1,*} (ORCID: 0009-0005-8255-9960)\\
\textbf{Klaas P. Pruessmann}\textsuperscript{1} (ORCID: 0000-0003-0009-8362) \\ \\
\noindent\textbf{1} Institute for Biomedical Engineering, ETH Zurich and University of Zurich, Switzerland\\ \\
\noindent\textbf{*} Corresponding author:\\ \\
\begin{tabular}{r l}
     \textbf{Name} & Dr. sc. Samuel Bianchi\\
     \textbf{Institute} & Institute for Biomedical Engineering  \\
     \textbf{Department} & Department of Information Technology and Electrical Engineering\\
     \textbf{University} & ETH Zürich \\ 
     \textbf{Address} & Gloriastrasse 35 \\
     & 8092 Zurich \\
     & Switzerland \\
     \textbf{E-mail} & bianchi@biomed.ee.ethz.ch
\end{tabular}
\newpage

\newpage

\section{Introduction}\label{sec_intro}
Blood oxygen level-dependent (BOLD) functional magnetic resonance imaging (fMRI, \cite{Ogawa1992}) has become one of the most frequently used functional neuroimaging techniques over the past three decades. Local neuronal activity causes an increased need for energy. Through neurovascular coupling, neuronal activity leads to changes in cerebral blood volume and flow, as well as the ratio of oxygenated to deoxygenated blood (\cite{Ogawa1993},\cite{Buxton1998}). These so-called hemodynamic responses can be detected by magnetic resonance imaging. Most pulse sequences used for fMRI are capable of generating an image in a time span of less than a second to a few seconds at most. For example, 2D or multiband EPI (echo planar imaging, \cite{Mansfield1977}, \cite{Larkman2001}), and 3D EPI (\cite{Song1994}, \cite{Mansfield1995}) are frequently implemented in practice. For 2D and multiband sequences, the object is segmented into slices, which are imaged individually. Multiple slices are imaged at once for multiband sequences; however, all data required to reconstruct an image of the excited slices are sampled after exactly one radio frequency (RF) excitation. Different slices are excited until all slices within the field-of-view (FOV) have been imaged. For 3D sequences, the full FOV is excited and the k-space gets segmented. After each RF excitation, one k-space segment is sampled. Hence, one image can be reconstructed after multiple RF excitations and the complete sampling of all k-space segments. With respect to timing, this imposes an important distinction between the two sequence categories. The acquisition time of each slice is known with great precision for 2D and multiband sequences. However, for 3D sequences, the timing of the image data is known with less precision because the k-space for a given image was sampled during multiple readouts. In fMRI, time series of many images are analyzed to infer on neuronal activity. However, a sophisticated account for the timing of 3D fMRI data is still lacking, and the implications for time series analysis are therefore unknown. In this document, a model for the timing of 3D fMRI data will be presented. In addition, we link the model and its properties to a large body of existing literature.\\
Before discussing the existing literature about 3D fMRI, we wish to quickly review the most relevant literature about 2D fMRI. Classically, general linear models (GLMs, \cite{Friston1994}) are used to infer on brain activity when analyzing fMRI time series. The GLM design matrix contains a set of regressors whose weighted sum provides a spatio-temporal model for the voxel time series. The regressors only depend on time and the respective weights only on coordinates in space. Importantly, it is possible to specify regressors, or equivalent model effects in the data, which exceed the band limit given by the repetition time ($T_R$) according to the Nyquist-Shannon sampling theorem (\cite{Friston1998}). This is important not only for modeling effects of interest but also for modeling confounding factors such as physiological noise. Physiological noise contains harmonic components up to $2Hz$ (or potentially higher) as, for example, shown in Figure 1 of \cite{Agrawal2020}. The generation of the regressors has to be done with care, as the slices of one image are temporally inconsistent. Three accounts for this inconsistency are known in the literature.
\begin{enumerate}
    \item Slice-timing correction using sinc-interpolation \cite{Henson1999}, \cite{Sladky2011}, \cite{Parker2017}, \cite{Parker2019}
    \item Use of a flexible basis-set to model hemodynamic responses that can account for slight timing variations, for instance by including the temporal derivative of the canonical HRF \cite{Henson1999}, \cite{Sladky2011}
    \item Per slice synchronization of regressors \cite{Henson1999}
\end{enumerate} Clearly, the timing of 2D fMRI time series is well understood and considered a resolved issue in the literature. This still holds for multiband fMRI data. In \cite{Feinberg2010}, this fact is explicitly noted as an advantage of multiband sequences over 3D sequences. It is stated that the segmentation of the k-space introduces additional unwanted perturbations to the fMRI time series in a potentially complicated way. In practice, the functional sensitivity of 3D sequences matches the sensitive to multiband sequences (\cite{Stirnberg2017}, \cite{LeSter2019}) when incorporating physiological noise correction. \\
The time series stability of 3D fMRI data has been investigated in terms of temporal signal-to-noise ratio (tSNR) in multiple studies. The tSNR is believed to be non-linearly related to the SNR of the image (\cite{Krueger2001a}, \cite{Krueger2001b}, \cite{Murphy2007}). The SNR is dependent on thermal and systematic noise and the tSNR is additionally dependent on physiological noise. 
\begin{equation}
    \text{tSNR} = \frac{\text{SNR}}{\sqrt{1 + \lambda^2 \cdot \text{SNR}^2}}
\end{equation}
$\lambda$ models the reduction of tSNR, compared to SNR, due to physiological noise sources. In \cite{vanderZwaag2012}, it is shown that for 3D sequences $\lambda$ increases linearly with the number of k-space segments. Physiological processes are suspected to be the cause of this effect. Earlier assessments of 3D sequences for fMRI already indicated a strong need for physiological noise correction (\cite{Poser2010}). \cite{Lutti2013} and \cite{Jorge2013} both showed that 3D sequences are more sensitive to physiological noise compared to 2D sequences. Incorporating physiological noise correction yielded an increase in tSNR in both studies. Furthermore, according to \cite{Lutti2013}, it is best practice to sample the regressors used for physiological noise correction at the time points when the temporally central k-space segment was sampled. Importantly, both physiological noise models included RETROICOR (\cite{Glover2000}). \cite{Tijssen2014}, in some sense, provides an early account of the timing of 3D fMRI data. The efficiency of RETROICOR was compared with the efficiency of RETROKCOR (\cite{Hu1995}). RETROICOR is an image-based physiological noise correction method, and is, therefore, subject to the temporal uncertainty of segmented 3D sequences. RETROKCOR is applied to k-space data for which the timing is known exactly. Surprisingly, it was found that RETROICOR performs as well as RETROKCOR and is preferable in practice due to its simplicity. It can be reasoned, based on \cite{Reynaud2017}, that RETROICOR plays a crucial role for physiological noise correction of segmented 3D fMRI data. The amount of variance modeled by RETROICOR increases drastically with the number of k-space segments, as shown in Figure 3 of \cite{Reynaud2017}. \\
The main topic of this document is a model for the timing of 3D fMRI data. The timing of 2D fMRI data is well understood and the timing of 3D data has not been modeled at all. We use the formulated model to address the following questions, which are based on the literature reviewed above. \\
\textbf{\textit{Can 3D fMRI data be used to infer on effects which exceed the band limit given by the Nyquist-Shannon sampling theorem?}} \\
Physiological noise contains harmonic components up to $2Hz$. The fact that physiological noise correction of 3D fMRI data with a volume repetition time of several seconds (e.g., $3.3s$ in \cite{Reynaud2017} is still effective suggests that this is actually the case. However, the underlying reason remains unclear. The effective acquisition duration can roughly be quantified as the number of k-space segments times $T_R$ (ignoring pauses between RF-excitations), referred to as the volume repetition time $T_{Vol}$ in the following. Hence, it is somewhat surprising that effects, which would require a much higher effective temporal resolution, can still be modeled. For 2D fMRI, it has been suggested that the use of regressors enables inference on effects which exceed the Nyquist-Shannon band limit (\cite{Friston1998}). However, in the 2D case, the effective acquisition duration of one slice is smaller or at most equal to $T_R$. This implies that, provided an accurate model of the voxel time series is available, the effective temporal resolution resulting from sequence timing of one slice becomes less of a limiting factor for inference.\\
\textbf{\textit{Can the decline of tSNR with increasing number of k-space segments be exclusively attributed to physiological processes?}} \\
Initially, it has been suspected that 3D fMRI data is highly sensitive to physiological noise. This suspicion has been confirmed by many studies. However, decreases in tSNR with an increasing number of k-space segments could also be explained by subject motion or instabilities of the MR scanner. Furthermore, it is important to note that different studies have used varying definitions of tSNR. In \cite{Poser2010} and \cite{vanderZwaag2012}, tSNR was calculated as the mean of the voxel time series divided by its standard deviation.
In contrast, \cite{Jorge2013} and \cite{Lutti2013} estimated tSNR as the ratio between the mean of the voxel time series and the standard deviation of the residuals obtained after regressing out physiological noise.\\
\textbf{\textit{Why is 3D fMRI so sensitive to physiological noise?}} \\
To the best of our knowledge, a clear answer to this question has not yet been found. Only the aforementioned argument by \cite{Feinberg2010} would provide an explanation. Still, there is no model that explains exactly how unwanted perturbations are precisely caused by the segmentation of the k-space and represented in the reconstructed image time series.  
 \\
\textbf{\textit{Why plays RETROICOR such a crucial role in the physiological noise correction of 3D fMRI data?}} \\
As noted previously, it is somewhat surprising that RETROICOR is as effective as RETROKCOR when applied to 3D fMRI data. Furthermore, it remains unclear why the relative amount of variance modeled by RETROICOR increases drastically with the number of k-space segments. In fact, these two observations appear to contradict the hypothesis that k-space segmentation introduces complex perturbations into the data. Why would RETROICOR remain so effective if the perturbations were complex? It is important to highlight that other components of the full physiological noise model used in \cite{Reynaud2017} do not show a similarly pronounced increase in the variance explained when the number of k-space segments increases. Specifically, the two other parts are based on cardiac and respiratory response functions (\cite{Chang2009}, \cite{Birn2008}), respectively. It appears that RETROICOR has truly unique properties when applied to 3D fMRI data. \\
\textbf{\textit{Why should regressors be sampled at time points coinciding with the acquisition of the temporally central k-space segment?}} \\ 
\cite{Lutti2013} empirically proved the effectiveness of this approach. Nonetheless, it is not clear why this is the case. 

\subsection{Outline}
Before describing the model, its building blocks, and its implications for data analysis, we provide a brief overview of the structure of this document. 
\begin{itemize}
    \item \myref{Section }{sec_transmag_model}: Introduces a simple yet versatile model for transverse magnetization based on the statistical description of a spin system's Larmor frequency. Crucially, two distinct time axes are introduced: one measuring the time since the last RF excitation and another measuring continuous time over the entire fMRI experiment.
    \item \myref{Section }{sec_bold_signal_enc_rec}: Describes BOLD signal changes based on the transverse magnetization model. Additionally, spatial encoding and image reconstruction are formally introduced and described.
    \item \myref{Section }{sec_isolating_timing_problem}: Introduces three key image representations: the actual image reconstructed from a segmented 3D acquisition, an optimal reference image unaffected by sequence timing, and an error image defined as their difference. Crucially, all three images are equally affected by any other source of artifacts.
    \item \myref{Section }{sec_spectral_timing_operators}: Defines three spectral operators in the frequency domain that generate the three key images, allowing the timing problem to be analyzed in terms of their mathematical properties. A simple proxy for the strenth of the timing error is also derived and evaluated.
    \item \myref{Section }{sec_total_discrete_timing_operators}: Defines three corresponding discrete-time operators. They allow to study the timing problem in terms of their singular value decomposition and to assess the stability of fMRI time series.
    \item \myref{Section }{sec_simulations}: Presents two instructive simulations using the total discrete timing operators: one illustrating how sequence timing leads to spatio-temporal artifacts, and another demonstrating its impact on image time series stability in practice.
    \item \myref{Section }{sec_timing_ts_analysis_GLM}: Examines the practical implications of these findings for fMRI data analysis using GLMs.
    \item \myref{Section }{sec_discussion}: Addresses the five core research questions we initailly formulated.
\end{itemize}

\newpage
\section{A simple but versatile model for the behavior of transverse magnetization}\label{sec_transmag_model}  
We begin by examining the transverse magnetization of a spin located at position $\mymat{r}$ and time $\tau$. The spin’s instantaneous Larmor frequency can be expressed as:
\begin{equation}
    f_{\text{Lam}}(\myvec{r},\tau) = \gamma B(\myvec{r},\tau)
\end{equation}
In this expression, $B(\myvec{r},\tau)$ refers to the magnetic field experienced by the spin on a molecular scale, and $\gamma$ denotes the gyromagnetic ratio. The resulting phase of the spin after a radio frequency (RF) excitation is described by:
\begin{equation}
    \varphi(\myvec{r},t,\tau) =  \int_{\tau - t}^{\tau}{f_{\text{Lam}}(\myvec{r},\tau')d\tau '} 
\end{equation}
In this formula, two different time axes have been introduced. $\tau$ measures the time over the entire fMRI experiment. $t$ measures the time that has passed since the last RF-excitation. In a pulsed MR experiment, these time axes are linked according to $t =\tau - \lfloor \tau/T_R \rfloor T_R$, with $T_R$ being the repetition time. However, we will treat them as separable in the presented framework (the benefits of doing so will become clearer later). Finally, the transverse magnetization of a single spin is given as: $e^{i\varphi(\myvec{r},t,\tau)}$. 
\\ At the macroscopic scale, the Larmor frequency of individual spins is no longer of interest, since each location contains a spin ensemble, composed of an extremely large number of spins. Therefore, it is more appropriate to examine the probability distribution of Larmor frequencies at a given position $\myvec{r}$ and time $\tau$. 
\begin{equation}
    p_{\text{Lam}}(f;\myvec{r},\tau) = p_{\text{Lam}}(f;\theta(\myvec{r},\tau)))
\end{equation}
The right-hand side of this equation implies that the probability distribution can be parameterized by a set of spatially and temporally varying parameters $\theta(\myvec{r},\tau)$. These parameters represent tissue properties of the imaged brain, which naturally depend on location $\myvec{r}$ and may evolve over time $\tau$. When both location and time are fixed, the observed spin ensemble exhibits static stochastic properties. In this theoretical scenario, the phase of each spin is given by:
\begin{equation}
    \varphi(\myvec{r},t,\tau) = t f_{\text{Lam}}(\myvec{r},\tau)
\end{equation}
The normalized macroscopic transverse magnetization is given by integrating over the transverse magnetization of all individual spins (\cite{Kiselev2018}). 
\begin{equation}\label{eq:normalized_transverse_mag}
    \alpha(t;\theta(\myvec{r},\tau)) = \int_{-\infty}^{\infty}{e^{itf}p_{\text{Lam}}(f;\theta(\myvec{r},\tau)) df}
\end{equation}
Both time axes can be linked to obtain a formula for the normalized macroscopic transverse magnetization observed in physical reality. 
\begin{equation}
    \alpha(\tau - \lfloor \tau/T_R \rfloor  T_R;\theta(\myvec{r},\tau)) =  \int_{-\infty}^{\infty}{e^{i(\tau - \lfloor \tau/T_R \rfloor T_R) f}p_{f_{Lam}}(f;\theta(\myvec{r},\tau)) df}
\end{equation}
It is important to note that this derivation has been greatly simplified by initially treating $t$ and $\tau$ as separate variables and only linking them later. Without this separation, it would have been necessary to handle random walks of Larmor frequencies over the interval $[\tau - t,\tau]$ and their corresponding integrals over this interval.\\
\myref{Equation }{eq:normalized_transverse_mag} implies that whatever effect alters the local distribution of Larmor frequencies $p_{\text{Lam}}$ has an impact on the behavior of transverse magnetization. Because brain function is dynamic, our model is time-dependent. Transverse relaxation in biological tissue rarely follows a simple mono-exponential decay \cite{Kiselev2018}. Instead, it is a complex process influenced by tissue properties like the vascular architecture, cerebral blood volume/flow (\cite{Uludag2009}, \cite{Kim2012}).  All of these quantities might be subsumed into $\theta$. Various phenomena are associated with slight deviations of $\theta$ from baseline. Hemodynamic responses, as observed in BOLD fMRI data, are such a phenomenon.\\
The unnormalized transverse magnetization, $\rho\left(\myvec{r},t; \myvec{\theta}(\myvec{r},\tau)\right)$, can be described as:
\begin{equation}
    \rho\left(\myvec{r},t; \myvec{\theta}(\myvec{r},\tau)\right)  = P(\myvec{r})
    \alpha(t;\theta(\myvec{r},\tau))
\end{equation}
$P(\myvec{r})$ is just the initial magnitude. 
\paragraph{Example: Observing a time-varying $T_2^*$-decay with static off-resonance } \mbox{}\\
In physical reality a time-varying $T_2^*$-decay with a static off-resonance can be described as:
\begin{equation}
    \alpha(\tau - \lfloor \tau/T_R \rfloor T_R;B_0(\myvec{r}),T_2^*(\myvec{r},\tau)) = e^{(\tau - \lfloor \tau/T_R \rfloor T_R)(i\gamma B_{0}(\myvec{r}) - 1/T_2^*(\myvec{r},\tau)) }
\end{equation}
The separation of $t$ and $\tau$ leads to:
\begin{equation}
     \quad \alpha(t;B_0(\myvec{r}),T_2^*(\myvec{r},\tau)) = e^{t(i\gamma B_{0}(\myvec{r}) - 1/T_2^*(\myvec{r},\tau))}
\end{equation}
The distribution of Larmor frequencies, according to \myref{Equation }{eq:normalized_transverse_mag}, which results in this normalized transverse magnetization, is given by the Lorentz distribution: 
\begin{equation}
    p_{\text{Lam}}(f;B_0(\myvec{r}),T_2^*(\myvec{r},\tau)) = \text{Lo}(f;\gamma B_{0}(\myvec{r}),1/T_2^*(\myvec{r},\tau)) \quad \text{Lo}(x;\alpha,\beta) = \frac{1}{\pi \beta \left( 1- \left( \frac{x - \alpha}{\beta} \right)^2 \right) }
\end{equation}
We hope that this example helps to foster a better understanding of the formalism we have developed so far. 
\newpage
\section{BOLD signal changes, spatial encoding and reconstruction}\label{sec_bold_signal_enc_rec}
Using the derived model for the transverse magnetization it is possible to model BOLD signal changes in a very generic way. We can simply understand observed BOLD signal changes as being caused by deviations of $\theta(\myvec{r},\tau)$ from its value at baseline $\overline{\theta}(\myvec{r})$. Formally the unnormalized transverse magnetization can, therefore, be separated as:
\begin{equation}\label{eq:object_as_sum}
\begin{split}
    \rho\left(\myvec{r},t;\overline{\myvec{\theta}}(\myvec{r})+ \Delta \myvec{\theta}(\myvec{r},\tau)\right) & = 
     P\left(\myvec{r}\right) \alpha\left(t;\overline{\myvec{\theta}}(\myvec{r})+ \Delta \myvec{\theta}  (\myvec{r},\tau)\right) 
     \\ & = P\left(\myvec{r}\right)  \left( A\left(t;\overline{\myvec{\theta}}(\myvec{r})\right) +
    a\left(t;\overline{\myvec{\theta}}(\myvec{r}),\Delta \myvec{\theta}  (\myvec{r},\tau)\right) 
    \right) \\
    & = \overline{\rho}(\myvec{r},t; \overline{\myvec{\theta}}(\myvec{r})) + \Delta \rho \left(\myvec{r},t;\overline{\myvec{\theta}}(\myvec{r}), \Delta \myvec{\theta}  (\myvec{r},\tau)\right) 
\end{split}
\end{equation}
As fMRI-based neuroscience focuses on deviations form baseline $\Delta \theta$ and accordingly $\Delta \rho$ is typically the quantity of interest. 
\\The spatial encoding of the object, resulting in the measured coil signal $\sigma_\lambda$, is modeled as:
\begin{equation}
\begin{gathered}\label{eq_timing_sigma}
    \tau_{(m,n,v)} = T_S + mT_D+nT_R + vT_{Vol} \\
    \sigma_{\lambda}(\tau_{(m,n,v)})= \int{\rho\left(\myvec{r},t = \tau_{(m,0,0)};\theta(\myvec{r},\tau_{(m,n,v)}) \right) \text{Enc}_{(n,\lambda)}(\myvec{r},\tau_{(m,0,0)}) dV}
\end{gathered}
\end{equation}
$T_S$ marks the starting point of the acquisition. $m\in[0,M-1]$ counts samples within one readout taken at interval $T_D$, the receiver dwell time.  $n\in[0,N-1]$ counts the RF excitations, or equivalently k-space segments, per image. $v\in[0,V-1]$ counts the images that, after reconstruction, form the image time series and $\lambda\in[0,\Lambda-1]$ indexes the individual coils. The encoding function $\text{Enc}_{(n,\lambda)}$ might depend on gradient fields or the coil's sensitivity profile (see \cite{Wilm2011}). If a standard 3D EPI trajectory is implemented and coil sensitivity profiles ($s_\lambda(\myvec{r})$) are included the encoding function will be given as:
\begin{equation}
    \text{Enc}_{(n,\lambda)}(\myvec{r},t) =  s_\lambda(\myvec{r}) e^{i\myvec{k}(t,n)\myvec{r}} \quad
    \myvec{k}(t,n) = \begin{pmatrix}
        k_x(t) \\ k_y(t) \\ k_z(n)
    \end{pmatrix}
\end{equation}
This equation indicates that a single $k_x\times k_y$ plane of k-space is sampled during each RF-cycle. The encoding then progresses through k-space planes along the $k_z$-coordinate during subsequent RF-cycles. \\ 
Linear image reconstruction remains the standard in fMRI, with GRAPPA \cite{Griswold2002} and SENSE \cite{Pruessmann1999} being two primary examples. A reconstruction matrix $\mymat{R}$, fully describes liner image reconstruction. Typically, $\mymat{R}$ is calculated based on the implemented encoding function.
\begin{equation}\label{eq_image_recon}
    \myvec{\varrho}_{l,v} = \sum_{(m,n,\lambda)}{\mymat{R}_{l,(m,n,\lambda)}\sigma_{\lambda}(\tau_{(m,n,v)})}  \Longleftrightarrow \myvec{\varrho}_v = \mymat{R} \myvec{\sigma}_v
\end{equation}
$l$ indexes voxels in image domain.  The notation on the right implies that the sampled data of all coils, for a given image in the image time series, indexed by $v$, can be written into a vector $\myvec{\sigma}_v$. Multiplying this vector by $\mymat{R}$ results in a vector $\myvec{\varrho}_v$ that contains the vectorized, reconstructed image. \\
According to \myref{Equation }{eq:object_as_sum}, the object can be separated into two summands. This property is preserved by spatial encoding and linear image reconstruction. 
\begin{equation}
    \sigma_\lambda(\tau_{(m,n,v)}) = \overline{\sigma}_\lambda(\tau_{(m,n,0)}) + \Delta \sigma_\lambda(\tau_{(m,n,v)}) \quad \myvec{\varrho}_{v} = \overline{\myvec{\varrho}} + \Delta \myvec{\varrho}_{v}
\end{equation}
Again, only the second summand relates to transverse magnetization deviating from its dynamics at baseline. Hence, the first summand is not of interest here and will be mostly omitted in the following.
\newpage
\section{Isolating the timing problem}\label{sec_isolating_timing_problem}
\myref{Equations }{eq_timing_sigma}, and \myref{}{eq_image_recon} combined imply that the evolution of $\theta$ during the acquisition of a single image introduces an additional source of artifacts that is unique to fMRI. In structural MRI, one is typically concerned with transverse magnetization dynamics after RF excitation leading to image contrast and/or artifacts. In that case, $\theta$ is assumed to remain constant over time.  \\
If $\tau$ would not progress during the acquisition of one image $\theta$, and thus $\Delta \theta$ would remain static. Furthermore, if $t$ could still progress, only the artifacts arising from the sequence timing would be eliminated. All other artifacts caused by transverse magnetization dynamics or reconstruction errors would still be present in the reconstructed image. This motivates the formulation of a purely theoretical coil signal for which the temporal dependence on $\tau$ is decoupled from the temporal dependence on $t$.
\begin{equation}\label{eq_continues_coil_signal}
    \Delta\sigma_{(m,n,\lambda)}^{(c)}(\tau)= \int{\Delta \rho\left(\myvec{r},T_S + mT_D;\overline{\theta}(\myvec{r}),\Delta\theta(\myvec{r},\tau ) \right) \text{Enc}_{(n,\lambda)}(\myvec{r},T_S + mT_D) dV} 
\end{equation}
The superscript $(c)$ indicates that $\Delta\sigma^{(c)}$ is continues in $\tau$. $\Delta\sigma^{(c)}$ is a vector filled with coil data that continuously evolves during the fMRI experiment. Note that the coil data, which is actually sampled in practice (see \myref{Equation }{eq_timing_sigma}), can be written as:
\begin{equation}
    \Delta\sigma_{\lambda}(\tau_{(m,n,v)}) = \Delta\sigma_{(m,n,\lambda)}^{(c)}(\tau_{(m,n,0)}+vT_{Vol})
\end{equation}
Even though $\Delta\sigma^{(c)}$ is continues in $\tau$ it can still be multiplied by $\mymat{R}$. The result of this multiplication is the time-continues image $\Delta \varrho^{(c)}$.
\begin{equation}
    \Delta \varrho_l^{(c)}(\tau) = \sum_{(m,n,\lambda)}{\mymat{R}_{l,(m,n,\lambda)}\Delta\sigma_{(m,n,\lambda)}^{(c)}(\tau)} 
    \Longleftrightarrow
    \Delta \varrho^{(c)}(\tau) = \mymat{R} \Delta\sigma^{(c)}(\tau)
\end{equation}
The time-continuous image is affected by transverse magnetization dynamics during an RF cycle and by artifacts introduced through inaccurate image reconstruction. However, it remains unaffected by sequence timing.
\\A timing optimal image, $\Delta \varrho_v^{(Opt)}$, can be formulated as
\begin{equation}\label{eq_timing_optimal_image_ts}
   \Delta\varrho_{v}^{(Opt)}(T') = \Delta \varrho^{(c)}(T'+vT_{Vol})  
\end{equation}
This image is optimal with respect to sequence timing, as it is acquired instantaneously on the time axis spanned by $\tau$ and is therefore not subject to the temporal uncertainty inherent to 3D fMRI. However, it is still influenced by the transverse magnetization dynamics that occurs during an RF cycle, since t, which represents the time within the RF cycle, progresses during the hypothetical acquisition of this optimal image.
The reference point $T'\in [0,T_{Vol}]$ can be freely chosen. Deviations from the optimal image can be expressed as:
\begin{equation}
    \Delta \myvec{\varrho}_{v}^{(\varepsilon)}(T') = \Delta \myvec{\varrho}_{v} - \Delta\varrho_{v}^{(Opt)}(T')
\end{equation}
We will refer to $\Delta \myvec{\varrho}_{v}^{(\varepsilon)}$ as the (timing-)error image. Additionally, the actual, imperfect image, consistent with \myref{Equation }{eq_image_recon}, can be expressed as:
\begin{equation}\label{eq:timing_real_object_2}
 \Delta\varrho_{l,v} = \sum_{(m,n,\lambda)}{\mymat{R}_{l,(m,n,\lambda)}\Delta\sigma_{(m,n,\lambda)}^{(c)}(\tau_{(m,n,0)}+  vT_{Vol})}  
\end{equation}

\newpage
\section{Spectral timing operators}\label{sec_spectral_timing_operators}
The relationship of $\Delta\varrho_{v}^{(c)}$ to $\Delta\varrho_{v}$, $\Delta\varrho_{v}^{(Opt)}$ and $\Delta\varrho_{v}^{(\epsilon)}$ can be explored in spectral domain. Operators relating $\Delta\varrho_{v}^{(c)}$ to the latter three images will be defined. These operators model the effect
of sequence timing on 3D fMRI data in the spectral domain. Therefore, they are called spectral timing operators. 
\subsection{The operators: $\hat{\mymat{H}}$, $\hat{\mymat{H}}^{(Opt)}$, $\hat{\mymat{H}}^{(\epsilon)}$}
The Fourier transforms of $\Delta\sigma^{(c)}$ and $\Delta \varrho^{(c)}$ are given as:
\begin{equation}\label{eq:timing_cont_object_f}
    \Delta\hat{\sigma}^{(c)}(f) = \int_{-\infty}^{\infty}{\Delta\sigma^{(c)}(\tau)e^{-2\pi i f t}dt} \quad
    \Delta \hat{\varrho}^{(c)}(f) = \mymat{R}\Delta\hat{\sigma}^{(c)}(f)
\end{equation}
Using the Fourier transformation again, \myref{Equation }{eq:timing_real_object_2} can be rewritten in spectral domain:  
\begin{equation}
    \Delta \hat{\varrho}_l (f) = \left(\sum_{(m,n,\lambda)}{ \mymat{R}_{l,(m,n,v)} e^{2 \pi i f \tau_{(m,n,0)}} \Delta \hat{\sigma}^{(c)}_{(m,n,\lambda)}(f)  } \right) * 
    \frac{1}{T_{Vol}}\sum_{v=-\infty}^{\infty}{\delta\left(f-v/T_{Vol}\right)} 
\end{equation}
Here, $*$ represents the convolution. Matrix-vector notation can, again, be used instead to rewrite the summation over $(m,n,\lambda)$.  
\begin{equation}\label{eq:timing_object_f_1}
\begin{gathered}
    \Delta\hat{\varrho}(f) = \left( \mymat{R} \left(\mymat{I}_{\Lambda}\otimes \hat{\mymat{D}}(f)\right) \Delta \hat{\sigma}^{(c)}(f) \right) * \frac{1}{T_{Vol}}\sum_{v=-\infty}^{\infty}{\delta\left(f-v/T_{Vol}\right)}
    \\
    \hat{\mymat{D}}_{(m,n),(m',n')}(f) = \begin{cases}
    e^{2 \pi i f \tau_{(m,n,0)}} & \text{if }(m,n) = (m',n') \\
    0 & \text{else}
    \end{cases}
\end{gathered}
\end{equation}
$\hat{\mymat{D}}(f)$ is a $MN\times MN$ diagonal matrix with $e^{2 \pi i f \tau_{(m,n,0)}}$ on the diagonal, and $\otimes$ represents the Kronecker product. $\mymat{I}_\Lambda$ is a $\Lambda \times \Lambda$ identity matrix. \\
The spectral timing operator $\hat{\mymat{H}}(f)$ can now be defined according to: 
\begin{equation}
    \mymat{R} \left(\mymat{I}_{\Lambda}\otimes \hat{\mymat{D}}(f)\right) = \hat{\mymat{H}}(f) \mymat{R} \Longleftrightarrow  \hat{\mymat{H}}(f)  = 
    \mymat{R} \left(\mymat{I}_{\Lambda}\otimes \hat{\mymat{D}}(f)\right) \mymat{R}^\dagger
\end{equation}
This definition is reasonable as \myref{Equation }{eq:timing_object_f_1} can finally be rewritten as:
\begin{equation}\label{eq:timing_operator_applied}
    \Delta\hat{\varrho}(f) = \left( \hat{\mymat{H}}(f)\Delta \hat{\varrho}^{(c)}(f)  \right) * \frac{1}{T_{Vol}}\sum_{v=-\infty}^{\infty}{\delta\left(f-v/T_{Vol}\right)}
\end{equation}
Hence, we have found an spectral operator $\hat{\mymat{H}}(f)$ which maps the continues image $\Delta \hat{\varrho}^{(c)}$ to the actually retrieved image $\Delta\hat{\varrho}$ in spectral domain. This formalism enables the analysis of the timing problem for 3D fMRI in an isolated manner and independently of the object.  \\ 
The spectral timing optimal operator for the optimal image can be defined accordingly. 
\begin{equation}
    \Delta\hat{\varrho}^{(Opt)}(f;T') = ( e^{2 \pi i f T'} \Delta \hat{\varrho}^{(c)}(f)) * \frac{1}{T_{Vol}}\sum_{v=-\infty}^{\infty}{\delta\left(f-v/T_{Vol}\right)} \quad
    \hat{\mymat{H}}^{(Opt)}(f;T') = e^{2\pi i f T'}\mymat{I}_L
\end{equation}
As $\Delta \hat{\varrho}^{(\varepsilon)}$ is the difference of $\Delta \hat{\varrho}$ and $\Delta \hat{\varrho}^{(Opt)}$, its spectral timing error operator can be defined as the difference of $\hat{\mymat{H}}(f)$ and $\hat{\mymat{H}}^{(Opt)}(f;T')$. 
\begin{equation}\label{eq:timing_error_operator}
    \hat{\mymat{H}}^{(\varepsilon)}(f;T') = \hat{\mymat{H}}(f) - \hat{\mymat{H}}^{(Opt)}(f;T') = \mymat{R}\left( \mymat{I}_{\Lambda} \otimes \left(\mymat{\hat{D}}(f) - e^{2 \pi i f T'}\mymat{I}_{MN}\right) \right)\mymat{R}^{\dagger}
\end{equation}
\begin{equation}
    \Delta \myvec{\hat{\varrho}}^{(\varepsilon)}(f;T') = \left( \hat{\mymat{H}}^{(\varepsilon)}(f;T')\Delta \hat{\varrho}^{(c)}(f)  \right) * \frac{1}{T_{Vol}}\sum_{v=-\infty}^{\infty}{\delta\left(f-v/T_{Vol}\right)}
\end{equation}
It has to be pointed out that all spectral timing operators take the form of spatio-temporal filters. Therefore, it can already be concluded that the timing problem leads to spatio-temporal artifacts. 

\subsubsection{A proxy for the strength of the timing error}\label{sec:timing_proxy_strength_error}

The spectral timing error operator can be used to derive a sequence- and object-independent proxy for the strength of the timing error. The timing error image $\Delta \hat{\varrho}^{(\varepsilon)}$ is equal to $0$ if the spectral timing error operator $\hat{\mymat{H}}^{(\varepsilon)}$ is equal to $0$ (see \myref{Equation }{eq:timing_error_operator}). This is, not exclusively, the case if: 
\begin{equation}
    \mymat{\hat{D}}(f) - e^{2 \pi i f T'}\mymat{I}_{MN} = \mymat{0}
\end{equation}
The „distance“ to this scenario can be measured using the Frobenius-norm. 
\begin{equation}\label{eq:porxy_timingporblem}
\begin{gathered}
    \left|\left|\mymat{\hat{D}}(f) - e^{2 \pi i f T'}\mymat{I}_{MN} \right|\right|_F^2 = 
    \sum_{(m,n)}{\left|e^{2 \pi i f \tau_{(m,n,0)}}-e^{2 \pi i f T'} \right|^2} = \\
        2MN  \left(1 -
    \frac{U_{M-1}(\cos(\pi f T_D))}{M}
    \frac{U_{N-1}(\cos(\pi f T_R))}{N}
    \cos\left(2 \pi f (T' - \bar{\tau}) \right)
    \right)\\
        \bar{\tau} = \frac{\tau_{(M-1,N-1,0)} + \tau_{(0,0,0)}}{2}
\end{gathered}
\end{equation}
The proof of this equation is given in \myref{Appendix }{timing_proofs_3}. $U_X$ denotes the $Xth$ Chebyshev polynomial of the second kind. Note that $\bar{\tau}$ marks the center of the interval between the first and last time points ($\tau_{(0,0,0)}$ and $\tau_{M-1,N-1,0}$) when samples are acquired for a single image.\\
As the scaling by $2MN$ only accounts for the number of samples acquired per image, it is reasonable to define the proxy for the timing error $\epsilon(f)$ as: 
\begin{equation}\label{eq:proxy_timingproblem_final}
    \epsilon(f) = \sqrt{1 -
    \frac{U_{M-1}(\cos(\pi f T_D))}{M}
    \frac{U_{N-1}(\cos(\pi f T_R))}{N}
    \cos\left(2 \pi f (T' - \bar{\tau}) \right)} \quad \in [0,\sqrt{2}]
\end{equation}
A plot of Chebyshev polynomials applied to the cosine is shown in \myref{Figure }{fig_chebyshev_poly} on the upper half. As these are symmetric functions it suffices to plot them on the interval $[0,\pi]$. They all start at $0$ for $\varphi = 0$ and reach $1$ ($X$ even) or $-1$ ($X$ odd) for $\varphi = \pi$. With increasing $X$, they oscillate with a higher frequency but remain closer to $0$, except for the border of the interval.
\\The function $\sqrt{1-U_{X-1}(\cos(\varphi))/X}$ is very similar to $\epsilon(f)$, however a bit easier to study. It is plotted on the lower half of \myref{Figure }{fig_chebyshev_poly}. It increases much faster from $0$ to a value around $1$, with increasing $\varphi$, when $X$ is large. When $\varphi=\pi$, it evaluates to either $0$ or $\sqrt{2}$ as $U_{X-1}(\cos(\varphi))$ is either equal to $1$ or $-1$. 
\begin{figure}
\centerline{\includegraphics[width=17cm]{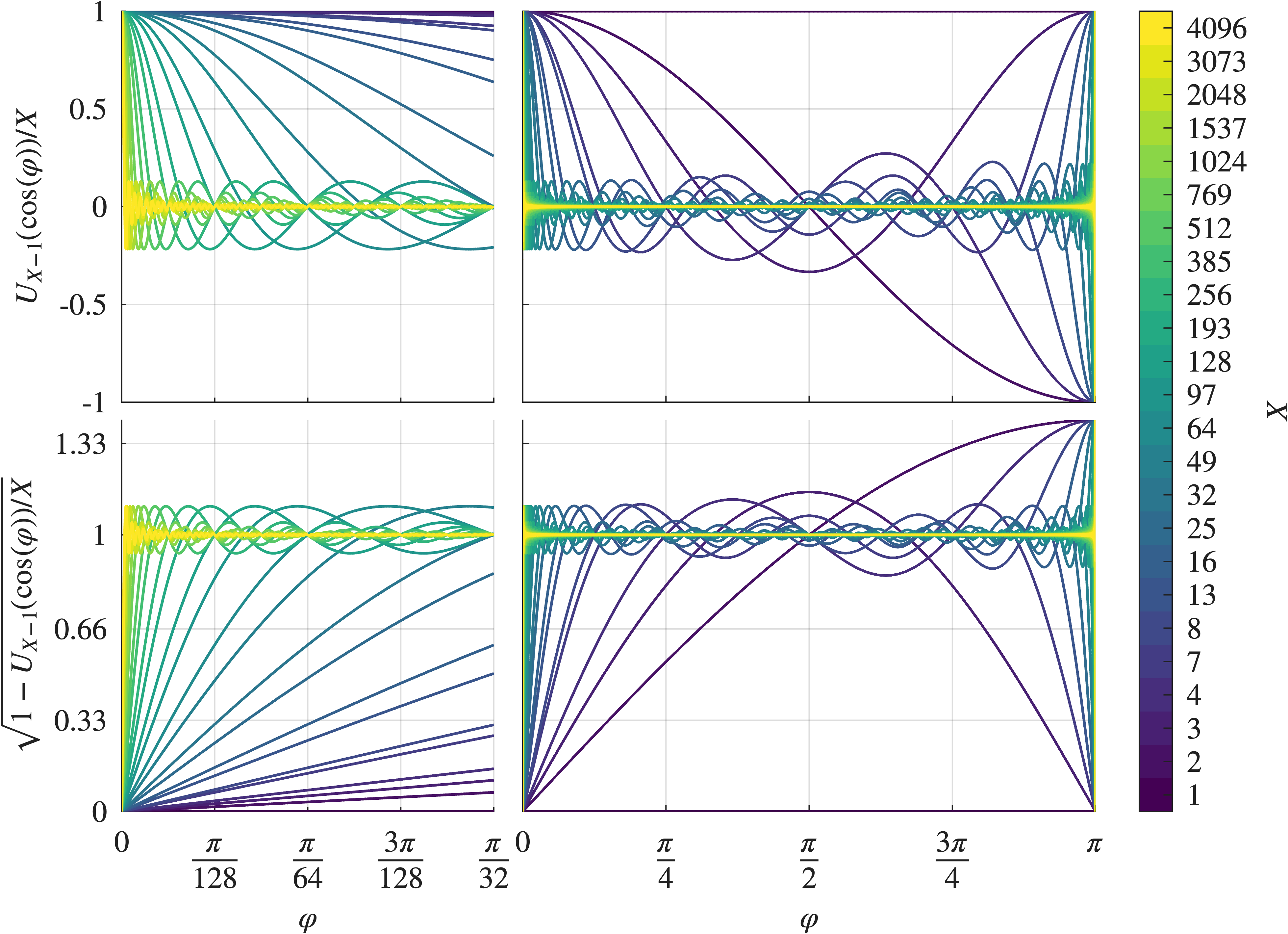}}
\caption{Top: Depiction of several Chebyshev polynomials of the second kind. The left plot provides a zoom-in, the right plot shows the polynomials over the full interval $[0,\pi]$. For $X$ even the polynomials are odd, for $X$ odd the polynomials are even. Bottom: $\sqrt{1-U_{X-1}(\cos(\varphi))/X}$ is fairly similar to $\epsilon(f)$ (see \myref{Equation }{eq:proxy_timingproblem_final}). It is, therefore, plotted to highlight the relationship to the original Chebyshev polynomial. \label{fig_chebyshev_poly}}
\end{figure}
\\Plots of the full proxy $\epsilon$ are shown in \myref{Figure }{fig_proxy_timing_error_MN}. $\varepsilon$ is always plotted against absolute frequency and no normalized quantity. This is due to the fact that hemodynamic responses as well as physiological noise fall onto relatively fixed, absolute frequency-intervals. Hence, it is reasonable to analyze $\epsilon$ on the absolute frequency axis as well. The only scenario where $M$ has any influence on the shown curves is observed for $N=1$. Hence, it can be concluded that it has a negligible influence on the proxy in practical scenarios where $N>1$. With increasing $N$, the proxy reaches the „saturation-region“ around $1$ faster. It can be shown that $\varepsilon(f) =1 $ occurs the first time at $f=1/T_{Vol} = 1/N T_R$, as this is the first time where $U_{N-1}(\cos(\pi f T_R))$ evaluates to $0$. Hence, the point where saturation is reached is only dependent on $N$ and $T_R$. We must conclude that $M$ as well as $T_D$ are not of great relevance here. 
\begin{figure}
\centerline{\includegraphics[width=17cm]{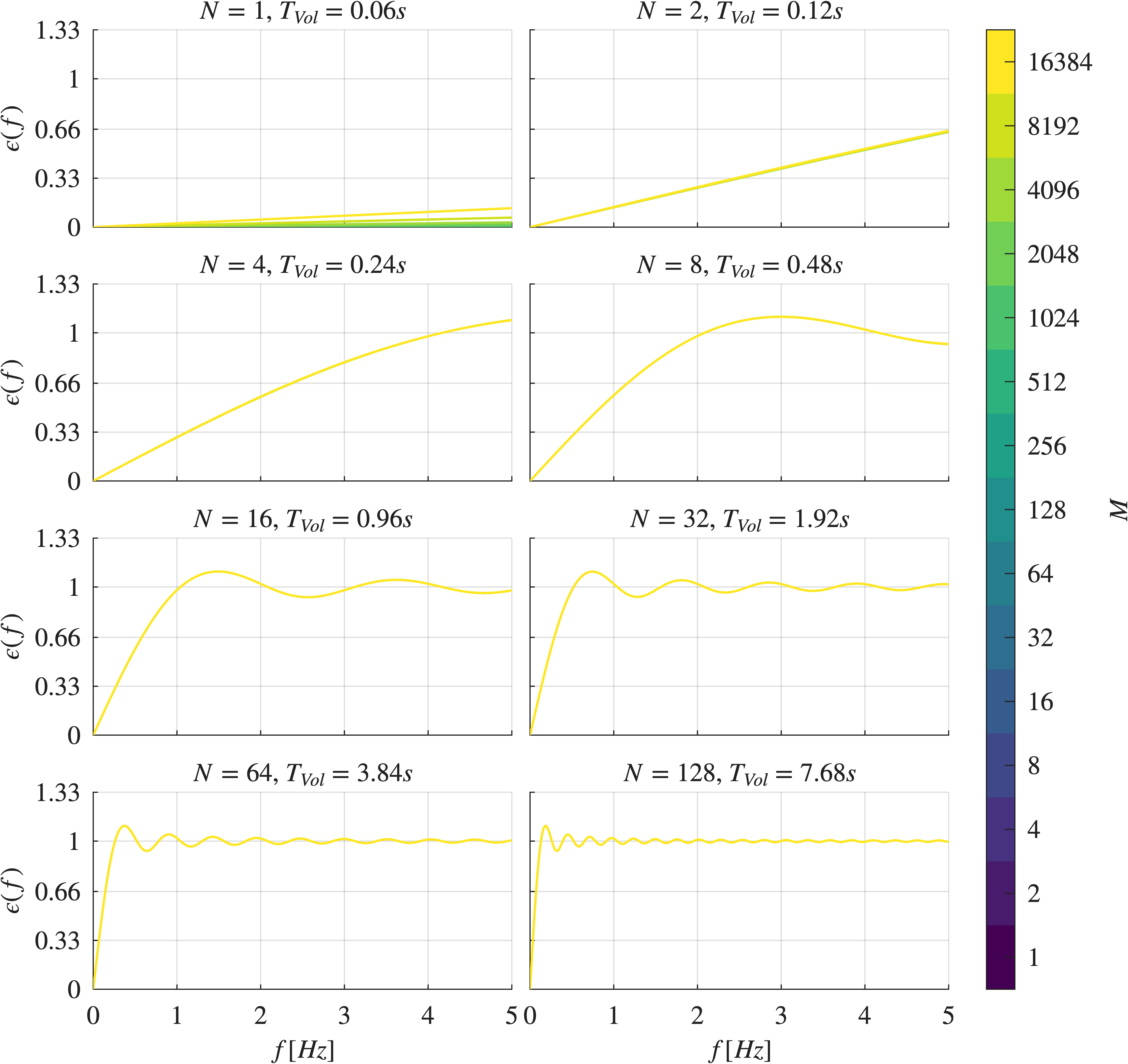}}
\caption{Plots of $\epsilon(f)$ (see \myref{Equation }{eq:proxy_timingproblem_final}) for various selections of $M$ and $N$. All other parameters were set to: $T_S=20ms$, $T_D =20ms/2^{14}$, $T_R=60ms$, $T' =\overline{\tau}$. In most cases all lines overlap. Therefore, only the yellow ($M=16384$) is visible.}\label{fig_proxy_timing_error_MN} 
\end{figure}
\\Considering real fMRI sequences the proxy predicts that the number of shots $N$ and repetition time $T_R$, with there product being equal to $T_{Vol}$, are the parameters that are strongly related to the potential level of the timing error. Intuitively, this makes a lot of sense. $M$ and $T_D$ are related to the within shot sequence timing where else $N$ and $T_R$ are related to the across shots sequence timing. Additionally the strength of the timing error approximately increases with the frequency of the signal sources causing the timing error. However, saturation is reached at $f=1/T_{Vol}$. 
\\So far the dependency of $\epsilon$ on the reference time point, $T'$, has not been addressed. Remember that $T'$ describes the shift of the sampling pattern of the timing optimal image time series ($\Delta \varrho_v^{(Opt)}$, see \myref{Equation }{eq_timing_optimal_image_ts}). Classical GLM anayses of 3D fMRI data do not offer an account for the timing problem. Hence, regressors are sampled on some temporal grid effectively modeling $\Delta \varrho_v^{(Opt)}$ and not $\Delta \varrho_v$. Therefore, the $T'$ that minimizes the strength of the timing error is also the one which should be used to generate the temporal grid on which regressors are sampled. Setting $T' = \overline{\tau}$, as already done for \myref{Figure }{fig_proxy_timing_error_MN}, minimizes the proxy for the strength of the timing error with respect to $T'$. Hence it is the optimal choice for minimizing the strength of the timing error as measured by $\epsilon$. \myref{Figure }{fig_proxy_timing_error_TP} depicts the dependency of $\epsilon$ on $T'$. For the optimal choice $T'=\overline{\tau}$, the proxy is minimized in the interval $[0,1/{T_{Vol}}]$. Due to the periodicity of the cosine term in $\epsilon$, other choices of $T'$ could theoretically minimize it in different intervals. However, this would always result in $|T'|>T_{Vol}$ which is nonsensical.
\begin{figure}
\centerline{\includegraphics[width=17cm]{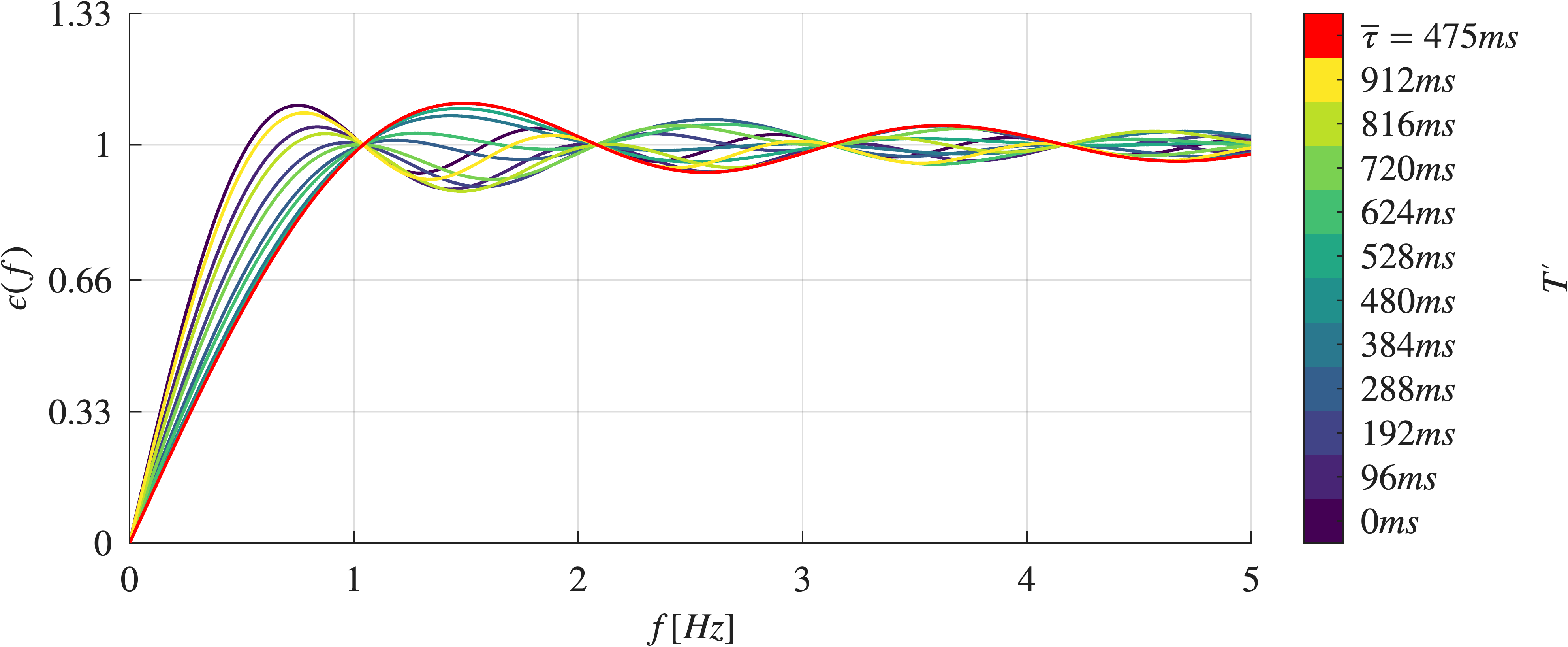}}
\caption{Plots of $\epsilon(f)$ (see \myref{Equation }{eq:proxy_timingproblem_final}) for various selections of $T'$. All other parameters were set to: $T_S=20ms$, $T_D =20ms/2^{14}$, $T_R=60ms$, $M=4096$, $N=16$. The proxy for the suggested, optimal choice of $T'$ is shown in red.}\label{fig_proxy_timing_error_TP}
\end{figure}

\newpage
\section{Total discrete timing operators}\label{sec_total_discrete_timing_operators}
Three additional timing operators can be defined in discrete time, rather than spectral-domain. To find them, we must consider the time-domain form of \myref{Equation }{eq:timing_operator_applied} and avoid matrix vector notion (for the moment).
\begin{equation}\label{eq:timing_operator_applied_time_domain}
    \Delta \varrho_{l,v} = \sum_{l'}{\sum_{(m,n,\lambda)}{\mymat{R}_{l,(m,n,\lambda)} \Delta\varrho_{l'}^{(c)}(\tau_{(m,n,0)}+vT_{Vol})\mymat{R}^{\dagger}_{(m,n,\lambda),l'}}}
\end{equation}
The proof is given in \myref{Appendix }{timing_proofs_1}. We can define a $L\times MN$ matrix
$\Delta\mymat{P}_v$ according to 
\begin{equation}\label{eq_def_delta_p_v}
    \Delta\mymat{P}^{(v)}_{l,(m,n)} = \Delta\varrho_{l}^{(c)}(\tau_{(m,n,0)}+vT_{Vol})
\end{equation}
$\Delta\mymat{P}^{(v)}$ contains all samples of $\Delta\varrho^{(c)}$ needed for the image with index $v$. 
Reintroducing matrix vector notation, \myref{Equation }{eq:timing_operator_applied_time_domain} can be written as
\begin{equation}
    \Delta \varrho_{v} = \mymat{R}\left(\mymat{R}^{\dagger} \circ \left(\myvec{1}_\Lambda \otimes (\Delta\mymat{P}^{(v)})^T \right)\right) \myvec{1}_L
\end{equation}
$\myvec{1}_\Lambda$ and $\myvec{1}_L$ are column vectors filled with $1$s of appropriate size, and $\circ$ represents the Hadamard product. Right-multiplying $\myvec{1}_L$ is equivalent to the summation over $l'$ and left-multiplying $\mymat{R}$ is equivalent to the summation over $(m,n,\lambda)$. This compact formulation is useful for simulation purpose. An other formulation, more useful for the theoretical analysis, can be found via the vectorization of $\Delta\mymat{P}^{(v)}$, denoted as $\text{vec}\left(\Delta\mymat{P}^{(v)}\right)$.
\begin{equation}\label{eq:timing_total_discrete_op_def}
    \Delta \varrho_{v} = \mymat{R}\left(\left( \myvec{1}_\Lambda  \otimes \mymat{I}_{MN}\right) \bullet \mymat{R}^\dagger\right) \text{vec}\left(\Delta\mymat{P}^{(v)}\right) = \mymat{\Omega}  \text{vec}\left(\Delta\mymat{P}^{(v)}\right)
\end{equation}
The proof of this equation is given in \myref{Appendix }{timing_proofs_2}, and $\bullet$ is the face-splitting product (a.k.a. row-wise Khatri-Rao product). We will refer to $\mymat{\Omega}$ (a $L\times LMN$ matrix) as the total discrete timing operator. This operator is total in the sense that it describes the total, or full, mapping of $\Delta\mymat{P}^{(v)}$ to $\Delta\varrho_v$, rather than just for a given frequency, and it is discrete as it operates on samples of $\Delta\varrho^{(c)}$. The face-splitting product essentially "stretches" $\mymat{R}^\dagger$ which results in the timing problem. 
\\For simplicity, we assume that we can find indices $m'$, and $n'$ such that
\begin{equation}
    T'\approx T_S + m'T_D+n'T_R
\end{equation}
The optimal total discrete timing operator $\mymat{\Omega}^{(Opt)}$ can be defined according to
\begin{equation}
    \Delta\varrho_v^{(Opt)} = \mymat{R}\left(\left( \myvec{1}_\Lambda  \otimes \myvec{1}_{MN}(\myvec{e}_{MN}')^T\right) \bullet \mymat{R}^\dagger\right) \text{vec}\left(\Delta\mymat{P}^{(v)}\right) = \mymat{\Omega}^{(Opt)}  \text{vec}\left(\Delta\mymat{P}^{(v)}\right)
\end{equation}
Here, $\myvec{e}_{MN}'$ is a unit column vector with $MN$ entries, and equal to $1$ at the index marked by $(m',n')$. It can easily be verified that
\begin{equation}\label{eq_omega_varepsiln_1_minus}
    \mymat{\Omega}^{(Opt)} = \left( \myvec{1}_\Lambda  \otimes \myvec{1}_{MN}(\myvec{e}_{MN}')^T\right) \bullet \mymat{I}_L \Longleftrightarrow
    \mymat{\Omega}^{(Opt)}_{l,(l',m,n)} =  
    \begin{cases}
        1 & \text{if } l = l' \text{ and } (m,n) = (m',n') \\
        0 & \text{else} 
    \end{cases}
\end{equation}
This essentially means that $\mymat{\Omega}^{(Opt)}$ selects the column of $\Delta\mymat{P}^{(v)}$ which is indexed by $(m',n')$. Finally the total discrete timing error operator is defined as:
\begin{equation}\label{eq:timing_total_discrete_error_operator}
\begin{gathered}
    \Delta\varrho_v^{(\varepsilon)} = \mymat{R}\left(\left( \myvec{1}_\Lambda  \otimes\left(\mymat{I}_{MN}  - \myvec{1}_{MN}(\myvec{e}_{MN}')^T\right)\right) \bullet \mymat{R}^\dagger\right) \text{vec}\left(\Delta\mymat{P}^{(v)}\right) = \mymat{\Omega}^{(\varepsilon)}  \text{vec}\left(\Delta\mymat{P}^{(v)}\right) \\
    \mymat{\Omega}^{(\varepsilon)}_{l,(l',m,n)} =  
    \begin{cases}
        \mymat{\Omega}_{l,(l',m,n)}-1 & \text{if } l = l' \text{ and } (m,n) = (m',n') \\
        \mymat{\Omega}_{l,(l',m,n)} & \text{else} 
    \end{cases}
\end{gathered}
\end{equation}
Hence, most entries of $\mymat{\Omega}^{(\varepsilon)}$ are equal to those of $\mymat{\Omega}$. \\ 
These matrix-vector formulations allow us to analyze the timing problem in terms of singular values and singular vectors. The singular values and singular vectors of $\mymat{\Omega}$ govern how much signal energy of $\Delta\mymat{P}^{(v)}$ is transferred to $\Delta \varrho_{v}$

\subsection{Sensitivity fields}
Before going into a SVD(singular value decomposition) based analysis of the total discrete timing operators, we wish to present a simpler approach of understanding them. Sensitivity fields (SFs) are reordered rows of $\mymat{\Omega}$. Formally, they can be defined as: 
\begin{equation}\label{eq:timing_def_sf}
    \text{SF}(l) = \text{vec}^{-1}\left((\mymat{\Omega}_{l,:})^H \right)
\end{equation}
The SF$(l)$ is an $L\times MN$ matrix, defined as the complex conjugate of the $l$th row of $\mymat{\Omega}$
reshaped such the its dimensions match the dimensions of $\Delta\mymat{P}^{(v)}$. The higher the correlation between $\Delta\mymat{P}^{(v)}$ with $\text{SF}(l)$, the higher the image intensity observed at $\Delta \myvec{\varrho}_{l,v}$. Each row of SF$(l)$, indexed by $l'$, can be interpreted as a template time series associated with the voxel of index $l'$. The greater the similarity between $\Delta \varrho^{(c)}_{l'}(\tau_{(m,n,0)}+vT_{vol})$ and $SF(l)_{l',(m,n)}$ during the acquisition of a given image (indexed by
$v$) the more $\Delta \varrho^{(c)}_{l'}(\tau_{(m,n,0)}+vT_{vol})$ will contribute to the reconstructed image at $\Delta \varrho_{l,v}$. Thus, the SF of voxel $l$ can be interpreted as a set of template time series for each voxel. The closer the actual intra image time series of voxel $l'$ aligns with the corresponding template, the greater the sensitivity of voxel $l$ to signals originating from voxel $l'$. Consequently, the temporal intra image dynamics of any voxel can influence the reconstructed signal of any other voxel. 
\\Examples of SFs are shown in \myref{Figure }{fig_timing_sense_field_ex}. For both examples, DFT-induced patterns are clearly visible. However, in the SENSE-case, this pattern are altered, as a result of the undersampling of k-space and the modulation by coil sensitivity profiles. Based on these examples it is already clear that $\mymat{\Omega}$ shows a strong dependence on $\mymat{R}$ which can easily be visualized and how $\mymat{\Omega}$ mixes signals originating from different voxels is not trivial. 
\begin{figure}
\centerline{\includegraphics[width=17cm]{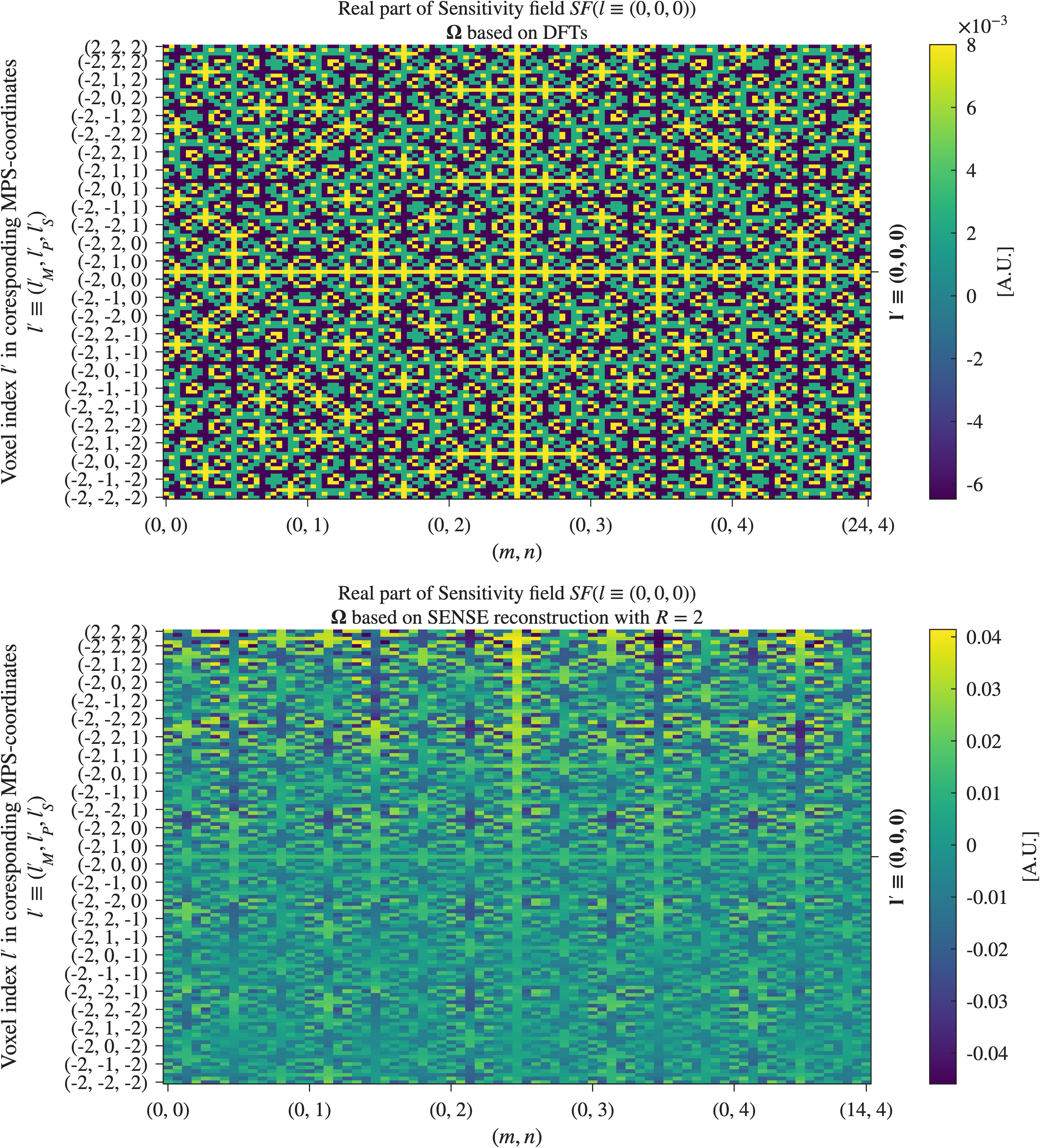}}
\caption{Real part of the sensitivity fields (SF) of the voxel in the center of the FOV (MPS-coordinates $(0,0,0)$). For both examples we chose, $L=5\times 5 \times 5$ and $N=5$. For the upper example $M=5\times 5$ was used and $\mymat{R}$ is an inverse 3D DFT without any undersampling. Therefore, a fully sampled 3D EPI acquisition consisting of $N=5$ k-space segments is simulated. For the lower example, $\mymat{R}$ represents a SENSE reconstruction represents a SENSE reconstruction including $\Gamma = 8$ coil sensitivity profiles and featuring and undersampling factor $R=2$ in primary phase encoding direction. Accordingly $M$ equals $5\times 3$. Everything else was kept consistent with the DFT-based case.}\label{fig_timing_sense_field_ex}
\end{figure}
\subsection{Mapping properties of the total discrete timing operators}\label{sec:timing_mapping_props_of_omega}
The matrix-vector formulations, as in \myref{Equation }{eq:timing_total_discrete_op_def}, allow us to analyze the timing problem in terms of singular values and singular vectors.  For brevity, most formulas are provided only for $\mymat{\Omega}$. However, analogous formulations can be obtained by substituting $\mymat{\Omega}$ with $\mymat{\Omega}^{(Opt)}$ or $\mymat{\Omega}^{(\varepsilon)}$.
\\Any matrix, including $\mymat{\Omega}$, can be expressed in terms of its singular value decomposition (SVD)
\begin{equation}
    \mymat{\Omega} = \mymat{U}\mymat{S}\mymat{V}^H
\end{equation}
Finding a tight relationship between the singular values of $\mymat{\Omega}$ and those of $\mymat{R}$ appears to be difficult. This can easily be seen by considering: 
\begin{equation}
    \mymat{\Omega}  \mymat{\Omega}^H= \mymat{U}  \mymat{S} \mymat{S}^H \mymat{U}^H = \mymat{R}( (\myvec{1}_\Lambda \myvec{1}^H_\Lambda \otimes  \mymat{I}_{MN})  \circ (\mymat{R}^\dagger (\mymat{R}^\dagger)^H ) )\mymat{R}^H    
\end{equation}
Multiplying by $\myvec{1}_\Lambda \myvec{1}^H_\Lambda \otimes  \mymat{I}_{MN}$ via the Hadamard product zeroes out many entries in $\mymat{R}^\dagger (\mymat{R}^\dagger)^H $. Importantly, without this step, the entire expression simplifies to $\mymat{I}_L$. However, this Hadamard product alters the mapping properties of $\mymat{\Omega}$ compared to those of $\mymat{R}$, such that it is not possible to find a direct relationship between their respective SVDs. The same holds for $\mymat{\Omega}^{(\varepsilon)}$. However, for $\mymat{\Omega}^{(Opt)}$ we find: 
\begin{equation}
    \mymat{\Omega}^{(Opt)} (\mymat{\Omega}^{(Opt)})^H = \mymat{I}_L
\end{equation}
Still, and regardless of the relationship between $\mymat{R}$ and $\mymat{\Omega}$ the mapping properties of $\mymat{\Omega}$ can be investigated. As we are interested in the temporal properties of $\Delta  \mymat{\varrho}$ over the full MR experiment (which means for all $v$), it is reasonable to define a $LMN \times V$ matrix $\Delta \tilde{\mymat{P}}$
\begin{equation}\label{eq:timing_full_continues_image}
    \Delta \tilde{\mymat{P}} = \left[\text{vec}\left(\Delta \mymat{P}_0\right), \dots,  \text{vec}\left(\Delta \mymat{P}_{V-1}\right)\right] 
    \quad
    \Delta\mymat{\varrho} = \mymat{\Omega} \Delta \tilde{\mymat{P}}
\end{equation}
Using $\Delta \tilde{\mymat{P}}$ we can express the inter-image covariance matrix of $\Delta\mymat{\varrho}$ as: 
\begin{equation}
    \Delta\mymat{\varrho}\Delta\mymat{\varrho}^H= \mymat{\Omega}  \Delta \tilde{\mymat{P}} \Delta \tilde{\mymat{P}}^H \mymat{\Omega}^H
\end{equation}
Following this idea, the temporal variance of $\Delta\mymat{\varrho}$ is given as: 
\begin{equation}
    \nu^{(Temp)}(\Delta \tilde{\mymat{P}},\mymat{\Omega}) = \frac{1}{V}\text{diag}\left( \mymat{\Omega}  \Delta \tilde{\mymat{P}} \Delta \tilde{\mymat{P}}^H \mymat{\Omega}^H \right) \quad \nu^{(Temp)}_l(\tilde{\mymat{P}},\mymat{\Omega}) = \frac{1}{V}\sum_{v}{|\Delta\mymat{\varrho}_{l,v}|^2} 
\end{equation}
The matrix $\frac{1}{V}\Delta \tilde{\mymat{P}} (\Delta \tilde{\mymat{P}} )^H$ describes the inter image covariance of the sampled continues image time series. Each element of this matrix is given as
\begin{equation}
    \frac{1}{V}\left(\Delta \tilde{\mymat{P}} (\Delta \tilde{\mymat{P}} )^H\right)_{(l,m,n),(l',m',n')} =  \frac{1}{V}\sum_{v}{\Delta\varrho_l(\tau_{(m,n,0)} + vT_{Vol})(\Delta\varrho_{l'}(\tau_{(m',n',0)} + vT_{Vol}))^*}
\end{equation}
The inter image covariance of the actual image time series with the continues image time series can also be described:open
\begin{equation}
     \frac{1}{V} \Delta \mymat{\varrho} ( \Delta\tilde{\mymat{P}} )^H = \frac{1}{V}\mymat{\Omega} \Delta \tilde{\mymat{P}} (\Delta \tilde{\mymat{P}})^H 
\end{equation}
This equation demonstrates that $\mymat{\Omega}$ maps the inter image covariance of the continuous image time series to the inter image covariance between the continues image time series and the image time series. Importantly, if $\frac{1}{V}\Delta \tilde{\mymat{P}}(\Delta \tilde{\mymat{P}} )^H = \mymat{I}_{MNL}$,
$\mymat{\Omega}$ directly represents the inter image covariance between the image and continuous image time series. Furthermore, under the same assumption of inter image independence of the continuous image time series, the inter image covariance of the image time series simplifies to $\mymat{\Omega} \mymat{\Omega}^H$.
Similarly, it becomes possible to assess the inter image covariance between the image time series and the optimal image time series, $\mymat{\Omega}(\mymat{\Omega}^{(Opt)})^H$, the error image time series with itself $\mymat{\Omega}^{(\varepsilon)}(\mymat{\Omega}^{(\varepsilon)})^H$, and the error image time series with the optimal image time series $\mymat{\Omega}^{(\varepsilon)}(\mymat{\Omega}^{(Opt)})^H$.
\\Apart from the temporal variance/covariance the total variance is also be formulated and investigated. 
\begin{equation}\label{eq:timing_total_variance}
\begin{gathered}
    \nu^{(Tot)}(\Delta \tilde{\mymat{P}}; \mymat{\Omega}) = \frac{1}{LV}\text{trace}\left( \mymat{\Omega} \Delta \tilde{\mymat{P}}(\Delta \tilde{\mymat{P}} )^H \mymat{\Omega}^H \right)   = \frac{1}{LV}\sum_{l,v}{|\Delta \varrho_{l,v}|^2}
\end{gathered}
\end{equation}
The total variance has a strong connection to the SVD of $\mymat{\Omega}$. Let $\mymat{W}$ be the representation of $\Delta \tilde{\mymat{P}}$ in the column-space of $\mymat{\Omega}$. 
\begin{equation}\label{eq_W_deltaP_tilde}
    \mymat{W} =\mymat{V}^H \tilde{\mymat{P}}
\end{equation}
Using $\mymat{W}$ it can easily be shown that
\begin{equation}\label{eq_vtot_svd_based}
    \nu^{(Tot)}(\Delta \tilde{\mymat{P}}; \mymat{\Omega}) = \frac{1}{LV} \sum_l{\left(\mymat{S}\mymat{S}^H\right)_{l,l}\left(\mymat{W}\mymat{W}^H\right)_{l,l}}
\end{equation}
Hence, the total variance depends on the representation of $\Delta \tilde{\mymat{P}}$ in the column-space of $\mymat{\Omega}$ and the singular values of $\mymat{\Omega}$. If we again consider the case where $\frac{1}{V}\Delta \tilde{\mymat{P}}(\Delta \tilde{\mymat{P}} )^H = \mymat{I}_{MNL}$ the total variance equals the trace of $\mymat{\Omega} \mymat{\Omega}^H$, or the sum over the squared singular values of $\mymat{\Omega}$.
\\The mapping properties of the total discrete timing operator(s) are closely related to the impact of sequence timing on tSNR. Using the derived formalism, tSNR can be expressed as:
\begin{equation}
    \text{tSNR}_l = \frac{|\overline{\varrho}_l|}{\sqrt{\frac{1}{V}\sum_v{|\Delta \varrho_{l,v}|^2}}}
\end{equation}
It is important to note that this definition of tSNR is based on complex-valued images, deviating from the conventional formulation. tSNR is inversely related to the temporal variance of $\Delta \varrho_{l,v}$. An increase in temporal variance leads to a decrease in tSNR. Since the primary goal here is to build intuition for the timing problem, it is sufficient to focus solely on the variance term. $|\overline{\varrho}_l|$ remains unaffected by the timing problem. This motivates to consider the following approximation:
\begin{equation}\label{eq_tsnr_approx}
    \frac{1}{L}\sum_l{\text{tSNR}_l^2} = \frac{1}{L} \sum_l{ \frac{|\overline{\varrho}_l|^2}{\frac{1}{V}\sum_v{|\Delta \varrho_{l,v}|^2}}} \approx \frac{\frac{1}{L} \sum_l{|\overline{\varrho}_l|^2} }{\nu^{(Tot)}(\Delta \tilde{\mymat{P}}; \mymat{\Omega})} \propto \frac{1}{\nu^{(Tot)}(\Delta \tilde{\mymat{P}}; \mymat{\Omega})}
\end{equation}
According to this approximation the mean squared tSNR is roughly inversely proportional to the total variance. Following this reasoning, and simply the definition in \myref{Equation }{eq:timing_total_variance}, the total variance can be considered a global measure for the stability of the image time series $\Delta \mymat{\varrho}$. Hence, we found a way to analyze the stability of image time series in terms of the SVD of the total discrete timing operator $\mymat{\Omega}$. Of course the same holds for $\Delta \mymat{\varrho}^{(Opt)}$ and $\Delta \mymat{\varrho}^{(\epsilon)}$ and the respective total discrete timing operators. Additionally, it must be emphasized that analyzing the total variance, as defined here, is more appropriate because it relates directly and solely to the timing problem. In contrast, the tSNR also depends on the mean image $\overline{\varrho}_l$, which is entirely unrelated to the timing problem. Therefore, the formalism presented here remains appropriate, even if the approximation in \myref{Equation }{eq_tsnr_approx} is not always completely accurate.

\subsection{DFT-based examples}\label{subsec_def_ex_sing_values_and_covar}
The squared singular values of $\mymat{\Omega}$  with $\mymat{R}$ representing a 3D DFT, are shown in \myref{Figure }{fig_singular_values_dft_example} for several choices of $N$, the number of RF-cycles required per image. We chose to display the squared singular values because the previously defined total variance depend on the squared singular values. They are all equal to $1$. The squared singular values of $\mymat{\Omega}^{(\varepsilon)}$ are all equal to $2$ except for the last one which is equal to $0$. Hence, the SVD of both operators remains highly structured when $\mymat{R}$ represents a 3D DFT. According to this analysis both operators show non to moderate tendencies to amplify the norm of vectors $\text{vec}(\mathbf{\Delta P}^{(v)})$ if they are well captures by a subspace included in $\mymat{V}$. 
\begin{figure}
\centerline{\includegraphics[width=17cm]{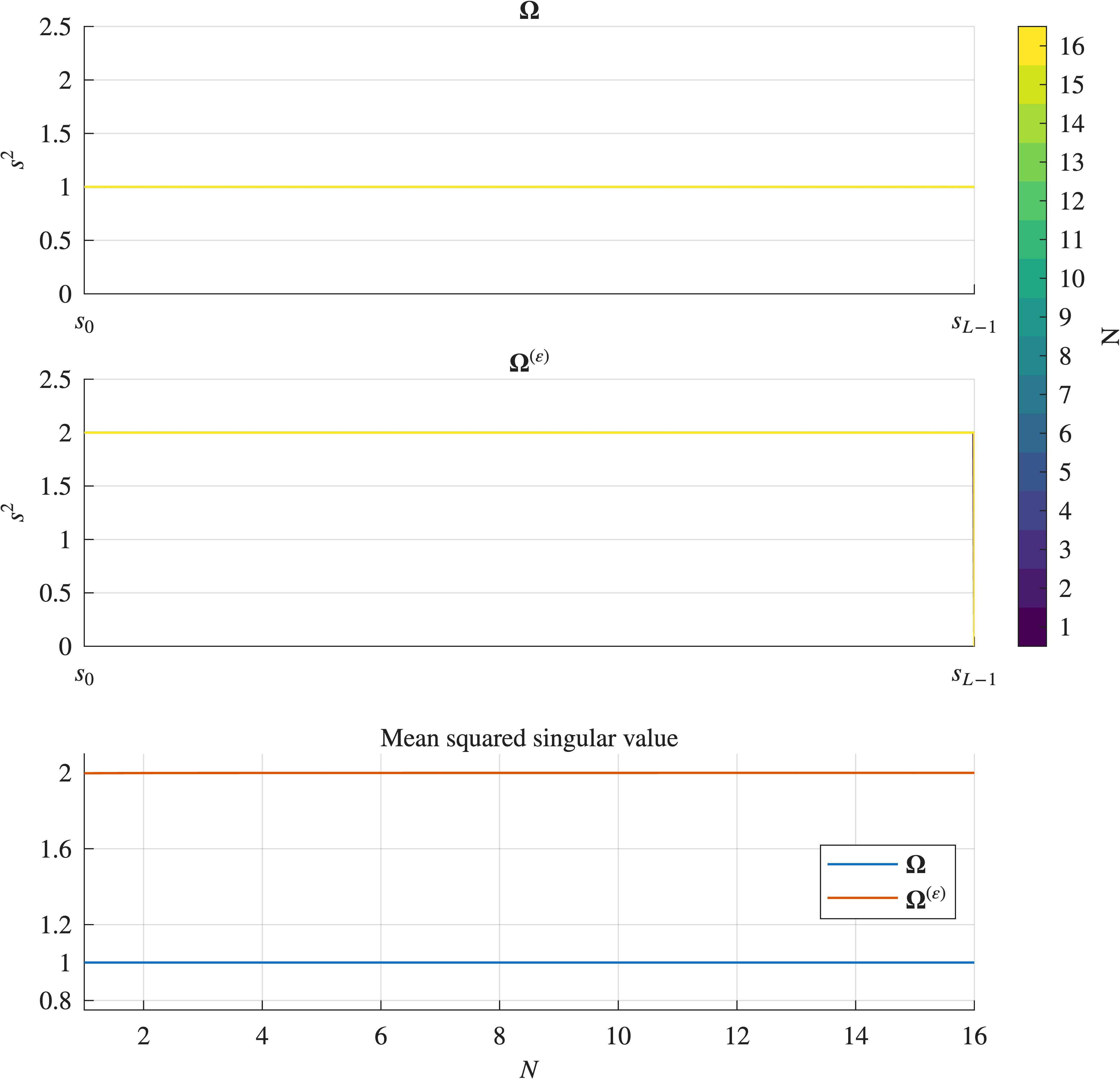}}
\caption{Top: Singular values squared of the total discrete timing operator $\mymat{\Omega}$ for multiple $N$ (all lines lie on top of each other). Middle: Singular values squared of the total discrete timing \textbf{error} operator $\mymat{\Omega}^{(\varepsilon)}$ for multiple $N$ (all lines lie on top of each other). Bottom: Mean of the squared singular values of both operators. For this example, $\mymat{R}$ is an inverse 3D DFT such that an $L=32\times32\times N$ image could be reconstructed. K-space was segmented in $N$ planes, each containing $M=32\times 32$ samples, always simulating a fully sampled 3D EPI. No coil sensitivity profiles were considered ($\Lambda = 1$). As both operators have dimensions $L\times LMN$, $L$ singular values are depicted starting with the largest one.}\label{fig_singular_values_dft_example}
\end{figure}
\\Parts of columns of $\mymat{V}$ are depicted in \myref{Figure }{fig_columns_V_dft_example}. Importantly, none of the columns of $\mymat{V}$ shows a spatiotemporal pattern reminiscent of natural objects, such as a human head. Therefore, it can be expected that the weights associated with a natural object ($\mymat{W}$, see \myref{Equation }{eq_W_deltaP_tilde}) are approximately uniformly distributed across the columns of $\mymat{V}$. Or, other words, all columns of $\mymat{V}$ are roughly equal in importance for capturing objects. By definition, $\mymat{\Omega}^{(\varepsilon)}$ is largely identical to $\mymat{\Omega}$, differing only by at few entries where $1$ is subtracted (see \myref{Equation }{eq_omega_varepsiln_1_minus}). Despite this seemingly minor adjustment, the effect on the column space of $\mymat{V}^{(\varepsilon)}$ is substantial. In $\mymat{V}^{(\varepsilon)}$, larger amplitude variations are observed. The phase patterns still exhibit features reminiscent of a 3D DFT, reflecting the underlying definition of $\mymat{R}$. Although the column space of $\mymat{V}^{(\varepsilon)}$ appears to be dissimilar to that of $\mymat{V}$, we still expect that all columns of $\mymat{V}^{(\varepsilon)}$ are roughly equally important for the representation of realistic objects. However, because nearly all squared singular values are close to $2$, the total variance of the error image time series $\mymat{\Delta \varrho}^{(\varepsilon)}$ is likely
greater than the total variance of the image time series $\mymat{\Delta \varrho}$ for most objects.
\begin{landscape}
\thispagestyle{empty}
\begin{figure}
\centerline{\includegraphics[width=27cm]{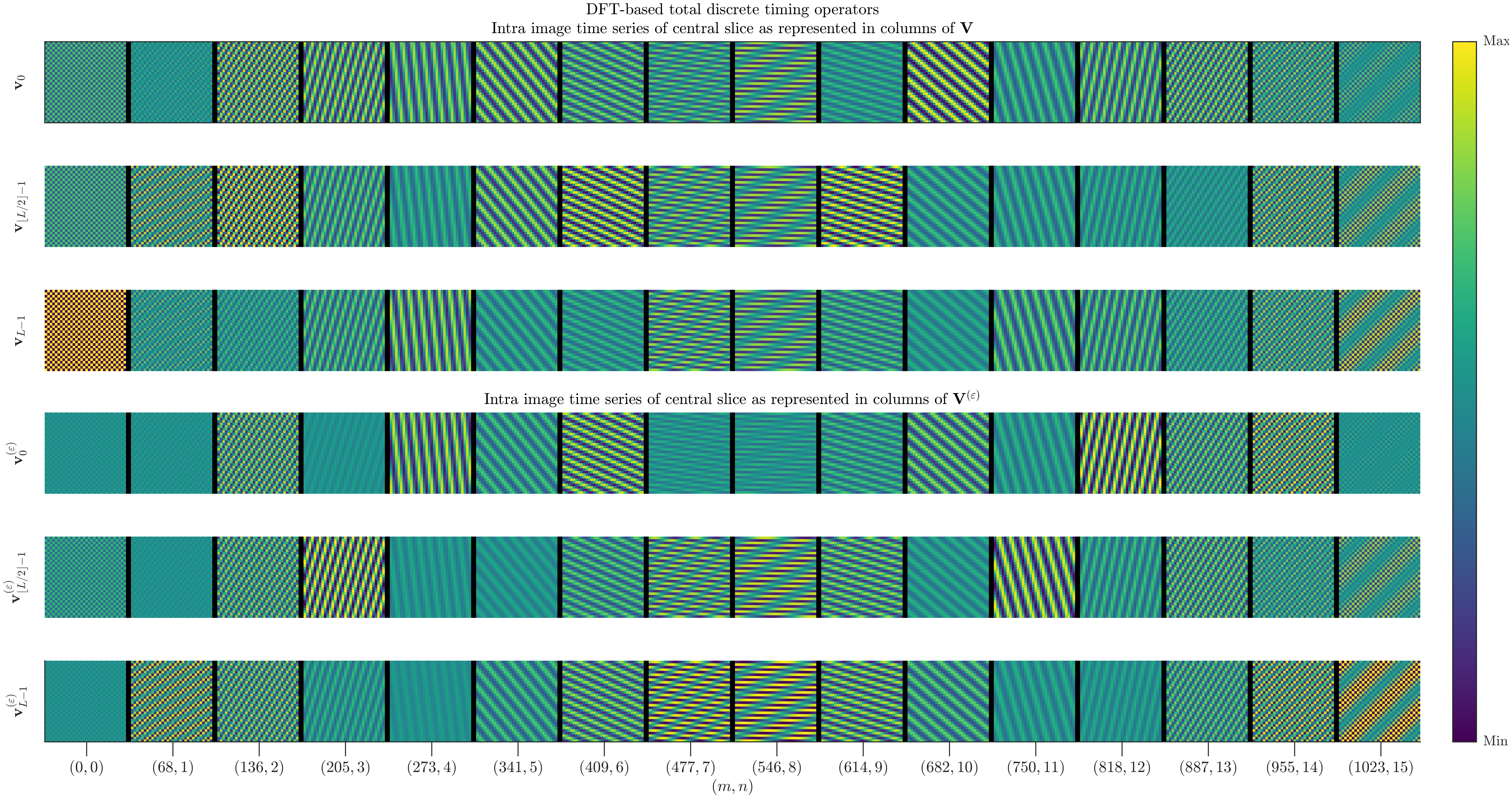}}
\caption{Real parts of columns ($\myvec{v}$,$\myvec{v}^{(\varepsilon)}$) of $\mymat{V}$ and $\mymat{V}^{(\varepsilon)}$ corresponding to the first, central, and last non-zero singular values (indices $0$, $\lfloor L/2\rfloor-1$, $L-1$) at selected intra-image time points $(m,n)$. Each column is reordered into an $L\times MN$ matrix analogous to $\mymat{\Delta P}^{(v)}$ (see \myref{Equation}{eq_def_delta_p_v}). Rows of this matrix referring to the central slice of the FOV have been reshaped to a slice and plotted above. For this example, $\mymat{R}$ is an inverse 3D DFT such that an $L=32\times32\times 16$ image could be reconstructed. K-space was segmented in $N$ planes, each containing $M=32\times 32$ samples, always simulating a fully sampled 3D EPI. No coil sensitivity profiles were considered ($\Lambda = 1$). As both operators have dimensions $L\times LMN$, $L$ singular values are depicted starting with the largest one.}\label{fig_columns_V_dft_example}
\end{figure}
\end{landscape}
\myref{Figure }{fig_covar_matrices_dft_example} displays various products of the total discrete timing operators. Under the assumption of inter-image independence of the sampled continuous image time series ($\frac{1}{V}\Delta \tilde{\mymat{P}}(\Delta \tilde{\mymat{P}})^H = \mymat{I}_{MNL}$), these products correspond to the covariance matrices between different image time series. Note that this represents the worst-case scenario for the timing problem, as the continuous image evolves so rapidly that no statistical dependence persists over a duration longer than $T_{Vol}$, making the actual image time series $\mymat{\Delta \varrho}$ highly erroneous. This effect is readily apparent in \myref{Figure }{fig_covar_matrices_dft_example}. Even though the product $|\mymat{\Omega} \mymat{\Omega}^H|$ still evaluates to $\mymat{I}_L$, the product $|\mymat{\Omega} (\mymat{\Omega}^{(Opt)})^H|$ reveals that $\mymat{\Delta \varrho}$ is barely correlated with $\mymat{\Delta \varrho}^{(Opt)}$ in this scenario. Furthermore, all voxels of $\mymat{\Delta \varrho}$ are equally correlated with all voxels of $\mymat{\Delta \varrho}^{(Opt)}$. Conversely, the timing error image time series $\mymat{\Delta \varrho}^{(\varepsilon)}$ becomes tightly correlated with $\mymat{\Delta \varrho}^{(Opt)}$, as shown by the product $\vert{}\mymat{\Omega}^{(\varepsilon)} (\mymat{\Omega}^{(Opt)})^H\vert{}$, and this correlation is spatially localized to the correct voxels.
\begin{figure}
\centerline{\includegraphics[width=17cm]{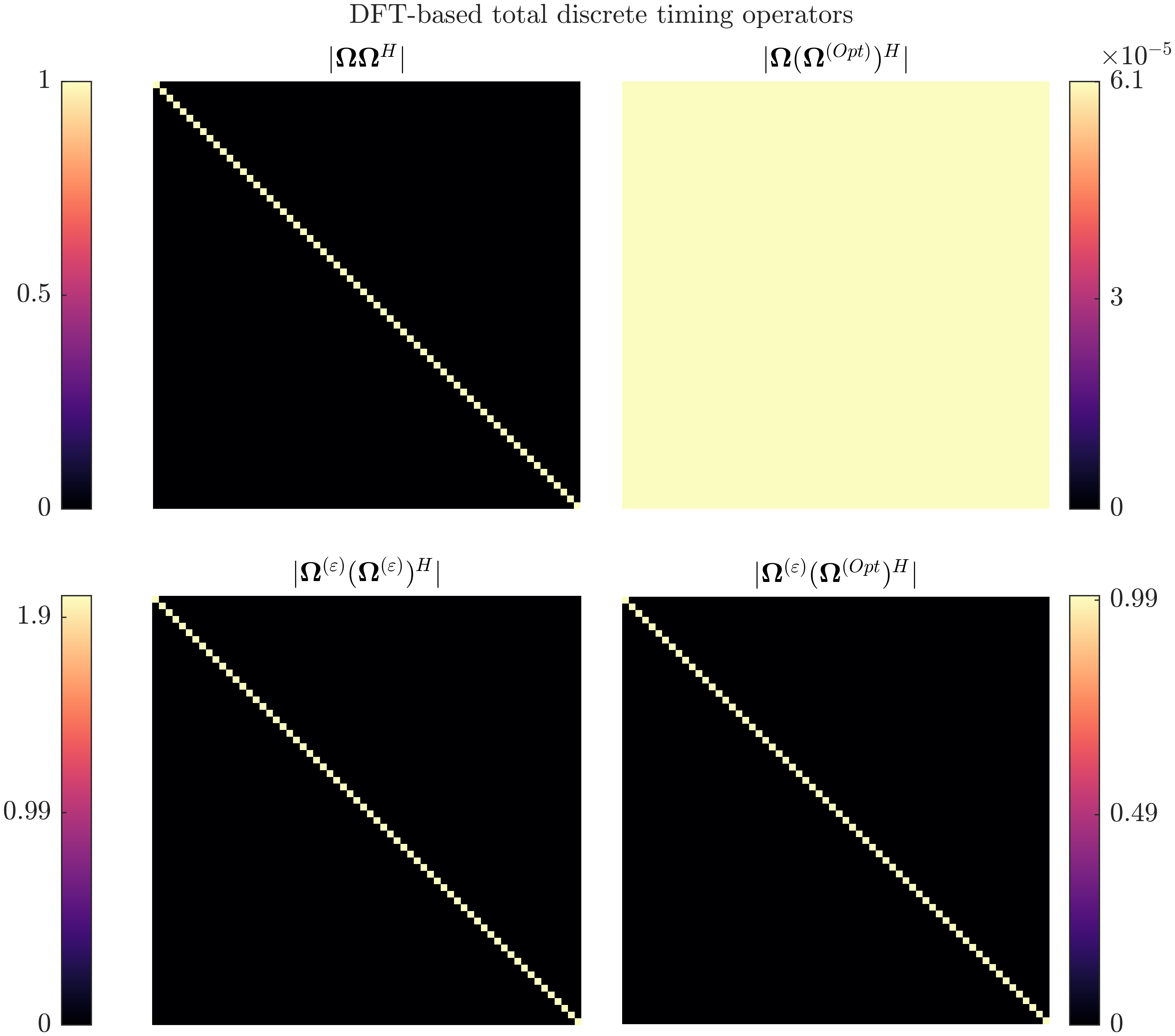}}
\caption{Magnitude of various inter image covariance matrices (indicated by $|\cdot|$) if inter image independence of the continues image time series is assumed ($\frac{1}{V}\Delta \tilde{\mymat{P}} (\Delta \tilde{\mymat{P}} )^H = \mymat{I}_{MNL}$) and scaling by $1/V$ is avoided. Top-Left: $\mymat{\Omega}\mymat{\Omega}^H=\Delta \varrho \Delta \varrho^H$, inter image covariance of the image. Top-Right: $\mymat{\Omega}(\mymat{\Omega}^{(Opt)})^H=\Delta \varrho (\Delta \varrho^{(Opt)})^H$, inter image covariance of the image and the optimal image. Bottom-Left: $\mymat{\Omega}^{(\varepsilon)}(\mymat{\Omega}^{(\varepsilon)})^H=\Delta \varrho^{(\varepsilon)} (\Delta \varrho^{(\varepsilon)})^H$, inter image covariance of the error image. Bottom-Right: $\mymat{\Omega}^{(\varepsilon)}(\mymat{\Omega}^{(Opt)})^H=\Delta \varrho^{(\varepsilon)} (\Delta \varrho^{(Opt)})^H$, inter image covariance of the error image and the optimal image. For this example, $\mymat{R}$ is an inverse 3D DFT such that an $L=32\times32\times 16$ image could be reconstructed. K-space was segmented in $N=16$ planes, each containing $M=32\times 32$ samples, simulating a fully sampled 3D EPI. No coil sensitivity profiles were considered ($\Lambda = 1$). For visualization purposes, all covariance matrices were down sampled by a factor of $8\times 8 \times 4$ in the respective spatial dimension.}\label{fig_covar_matrices_dft_example}
\end{figure}

\subsection{SENSE-based examples}
The squared singular values of $\mymat{\Omega}$ for the case where $\mymat{R}$ represents a SENSE reconstruction (see \cite{Pruessmann1999}) are shown in \myref{Figure }{fig_singular_values_sense_example}. Compared to the DFT case, larger singular values are observed. Nevertheless, most squared singular values remain smaller than the mean and below $1$. The spatiotemporal patterns of the columns of $\mymat{V}$ still appear very different from natural objects (see \myref{Figure }{fig_columns_V_sense_example}). However, unlike in the DFT case, the magnitude now exhibits additional spatial modulations, caused by the account for coil sensitivity profiles in the computation of $\mymat{R}$.  For $\mymat{\Omega}^{(\varepsilon)}$, the singular values increase further (see Figure \myref{Figure }{fig_singular_values_sense_example}), often exceeding $1$. Again, the columns of $\mymat{V}^{(\varepsilon)}$ are substantially altered by the subtraction of $\mymat{\Omega}^{(Opt)}$.
\begin{figure}
\centerline{\includegraphics[width=17cm]{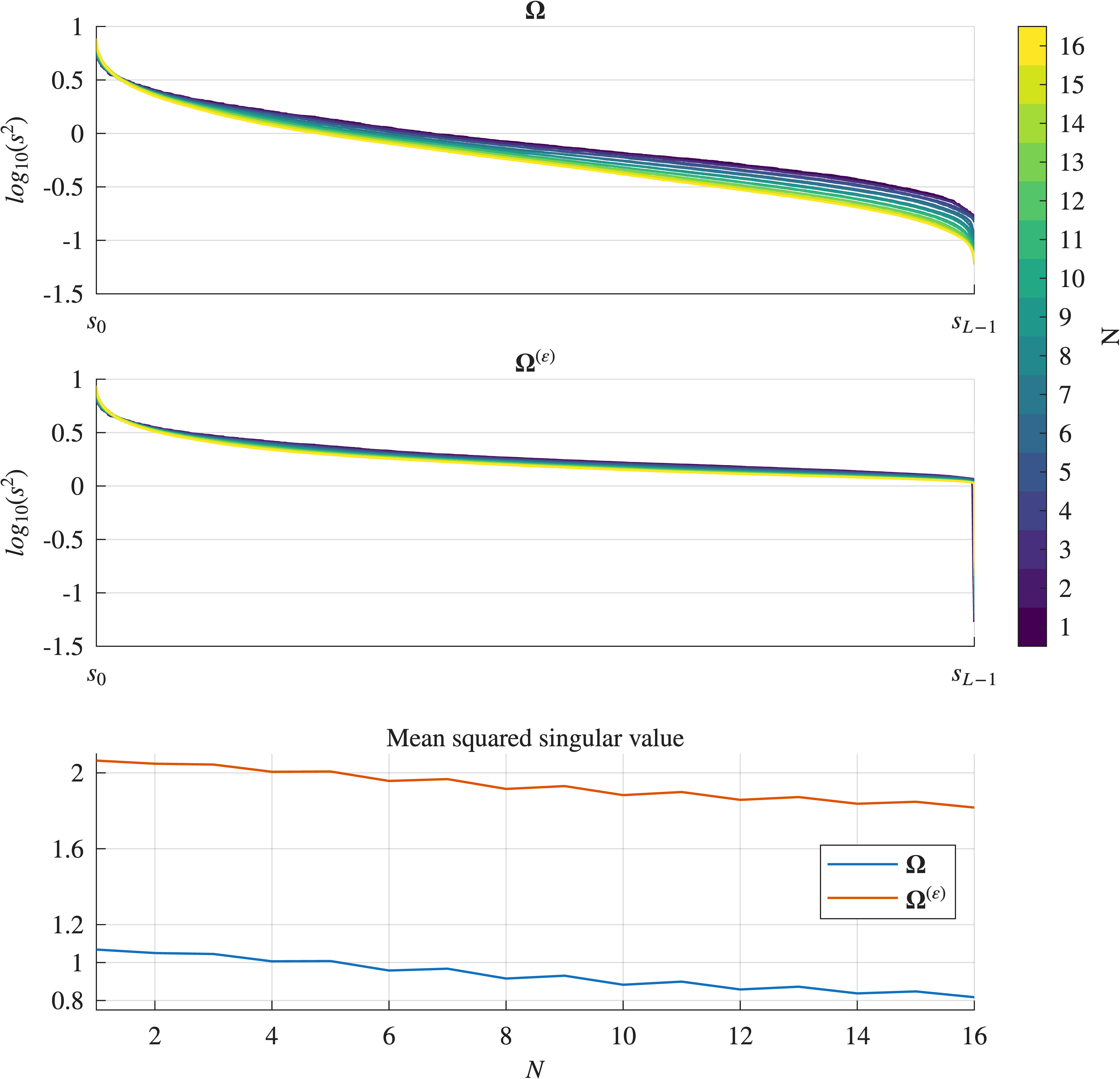}}
\caption{Top: Singular values squared of the total discrete timing operator $\mymat{\Omega}$ for multiple $N$ (all lines lie on top of each other). Middle: Singular values squared of the total discrete timing \textbf{error} operator $\mymat{\Omega}^{(\varepsilon)}$ for multiple $N$ (all lines lie on top of each other). Bottom: Mean of the squared singular values of both operators. For this example, $\mymat{R}$ represents a SENSE reconstruction (undersampling factor $R=2$ in primary phase encoding direction) such that an $L=32\times32\times 16$ image could be reconstructed. K-space was segmented in $N$ planes, each containing $M=32\times 16$ samples, simulating an undersampled 3D EPI. Coil sensitivity profiles of $\Lambda=8$ coils were considered. As both operators have dimensions $L\times LMN$, $L$ singular values are depicted starting with the largest one.}
\label{fig_singular_values_sense_example}
\end{figure}
\begin{landscape}
\thispagestyle{empty}
\begin{figure}
\centerline{\includegraphics[width=27cm]{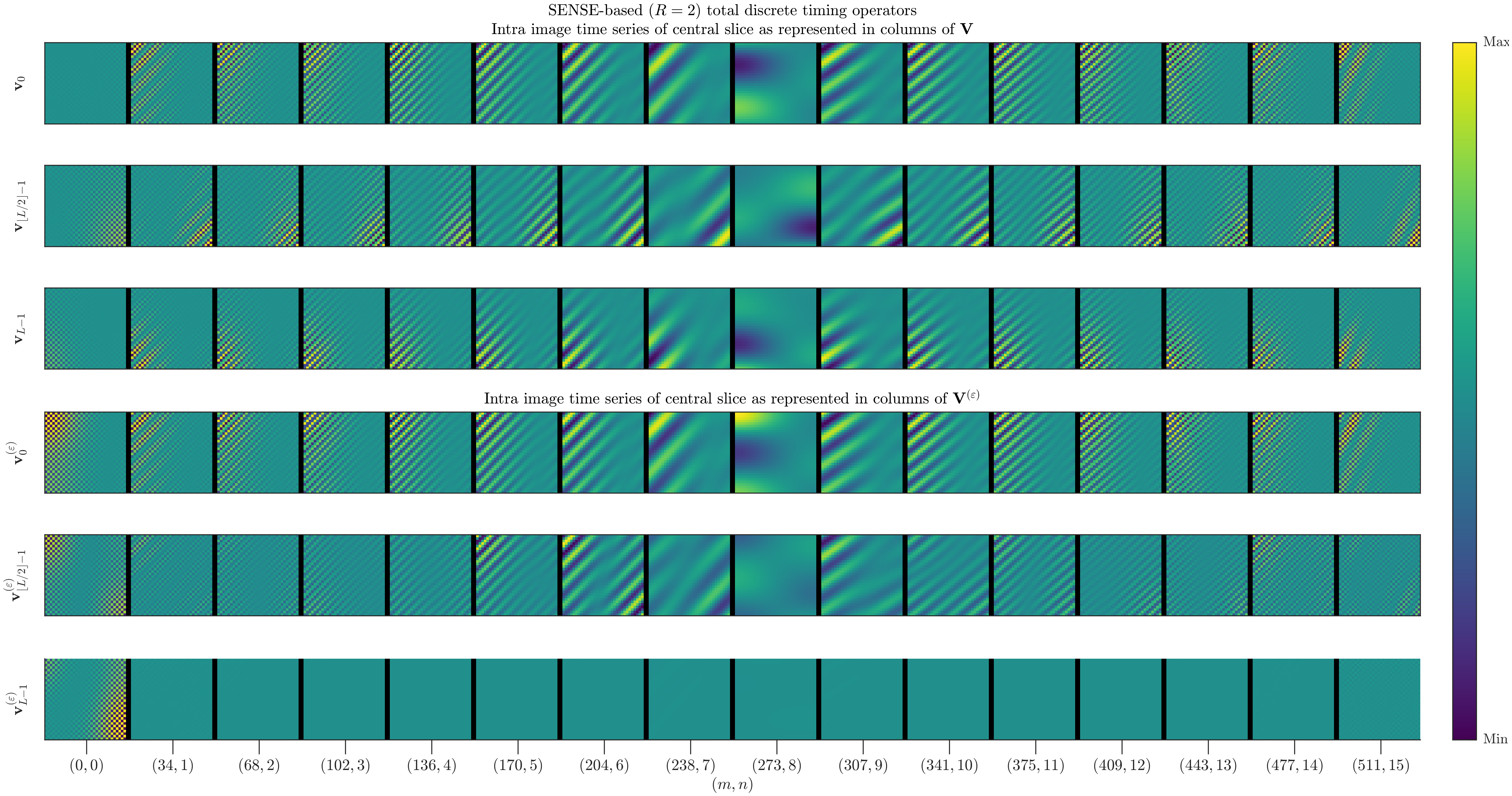}}
\caption{Real parts of columns ($\myvec{v}$,$\myvec{v}^{(\varepsilon)}$) of $\mymat{V}$ and $\mymat{V}^{(\varepsilon)}$ corresponding to the first, central, and last non-zero singular values (indices $0$, $\lfloor L/2\rfloor-1$, $L-1$) at selected intra-image time points $(m,n)$. Each column is reordered into an $L\times MN$ matrix analogous to $\mymat{\Delta P}^{(v)}$ (see \myref{Equation}{eq_def_delta_p_v}). Rows of this matrix referring to the central slice of the FOV have been reshaped to a slice and plotted above. For this example, $\mymat{R}$ represents a SENSE reconstruction (undersampling factor $R=2$ in primary phase encoding direction) such that an $L=32\times32\times 16$ image could be reconstructed. K-space was segmented in $N=16$ planes, each containing $M=32\times 16$ samples, simulating an undersampled 3D EPI. Coil sensitivity profiles of $\Lambda=8$ coils were considered.}
\label{fig_columns_V_sense_example}
\end{figure}
\end{landscape}
The covariance matrices depicted in \myref{Figure }{fig_covar_matrices_sense_example} look qualitatively similar to those shown in the DFT example. However, in this case, the diagonal elements are modulated and a few off-diagonal elements appear. Recall that $\frac{1}{V}\Delta \tilde{\mymat{P}} (\Delta \tilde{\mymat{P}} )^H = \mymat{I}_{MNL}$ holds, for example, when $\Delta \tilde{\mymat{P}}$ consists of sampled white noise. It is known that for SENSE reconstruction, the noise level varies across voxels and noise correlations between voxels can arise (as described in \cite{Pruessmann1999}). These characteristics directly explain the modulation of the diagonal elements and the presence of off-diagonal elements. Thus, the choice of the reconstruction matrix $\mymat{R}$ can have a pronounced impact on $\mymat{\Omega}$, $\mymat{\Omega}^{(\varepsilon)}$ and related quantities. This becomes further evident by considering \myref{Figure }{fig_totSENSE_vs_var}. Clearly, the summed squared coil sensitivity profiles appear to be correlated with the diagonal of $|\mymat{\Omega}\mymat{\Omega}^H|$. In areas of the FOV where the "total" coil sensitivity is low, the temporal variance, given by the diagonal of $|\mymat{\Omega}\mymat{\Omega}^H|$, appears to be high. Both quantities appear to be substantially correlated. 
\begin{figure}
\centerline{\includegraphics[width=17cm]{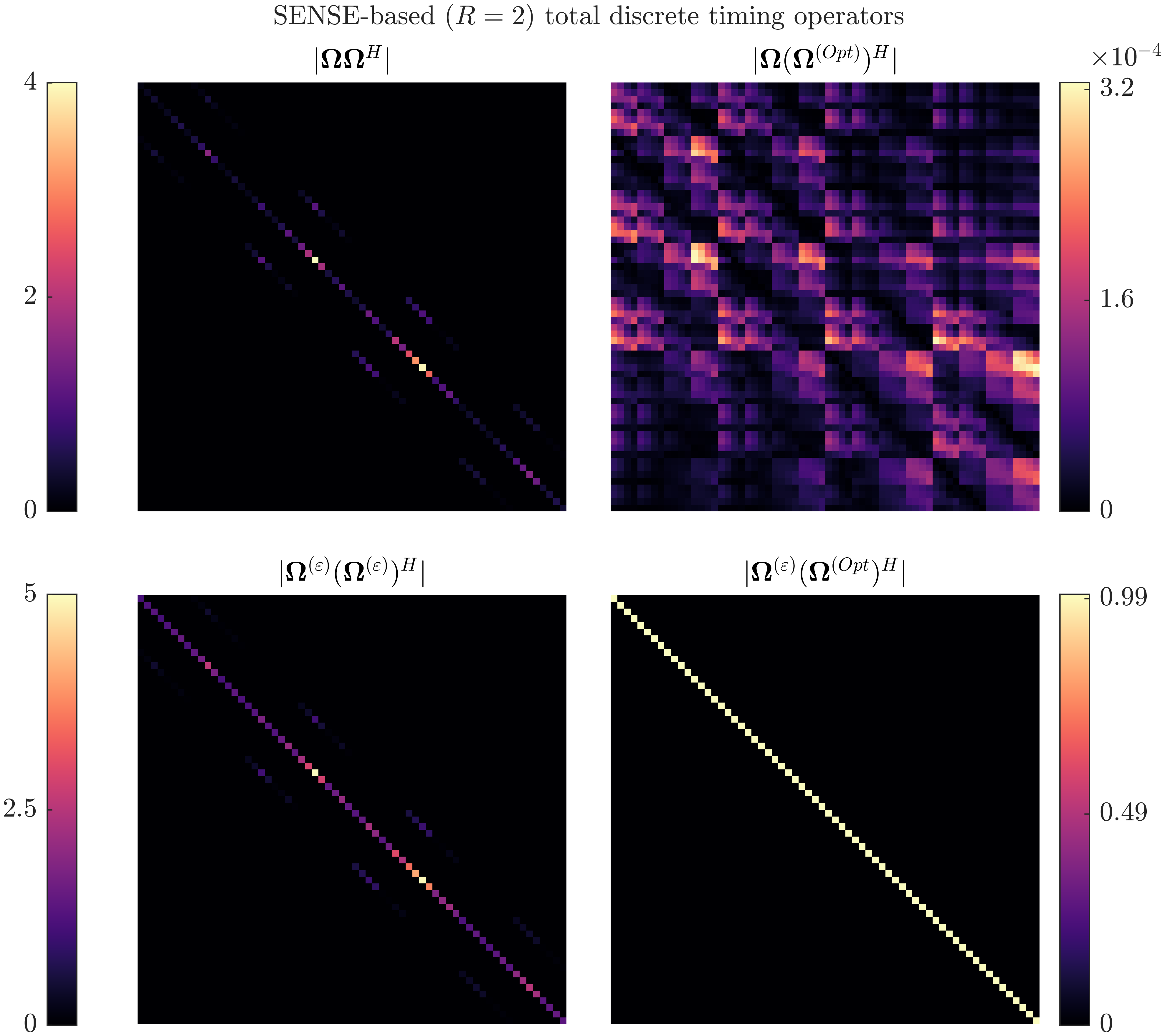}}
\caption{Magnitude of various inter image covariance matrices (indicated by $|\cdot|$) if inter image independence of the continues image time series is assumed ($\frac{1}{V}\Delta \tilde{\mymat{P}} (\Delta \tilde{\mymat{P}} )^H = \mymat{I}_{MNL}$) and scaling by $1/V$ is avoided. Top-Left: $\mymat{\Omega}\mymat{\Omega}^H=\Delta \varrho \Delta \varrho^H$, inter image covariance of the image. Top-Right: $\mymat{\Omega}(\mymat{\Omega}^{(Opt)})^H=\Delta \varrho (\Delta \varrho^{(Opt)})^H$, inter image covariance of the image and the optimal image. Bottom-Left: $\mymat{\Omega}^{(\varepsilon)}(\mymat{\Omega}^{(\varepsilon)})^H=\Delta \varrho^{(\varepsilon)} (\Delta \varrho^{(\varepsilon)})^H$, inter image covariance of the error image. Bottom-Right: $\mymat{\Omega}^{(\varepsilon)}(\mymat{\Omega}^{(Opt)})^H=\Delta \varrho^{(\varepsilon)} (\Delta \varrho^{(Opt)})^H$, inter image covariance of the error image and the optimal image. For this example, $\mymat{R}$ represents a SENSE reconstruction (undersampling factor $R=2$ in primary phase encoding direction) such that an $L=32\times32\times 16$ image could be reconstructed. K-space was segmented in $N=16$ planes, each containing $M=32\times 16$ samples, simulating an undersampled 3D EPI. Coil sensitivity profiles of $\Lambda=8$ coils were considered. For visualization purposes, all covariance matrices were down sampled by a factor of $8\times 8 \times 4$ in the respective spatial dimension.}
\label{fig_covar_matrices_sense_example}
\end{figure}

\begin{figure}
\centerline{\includegraphics[width=17cm]{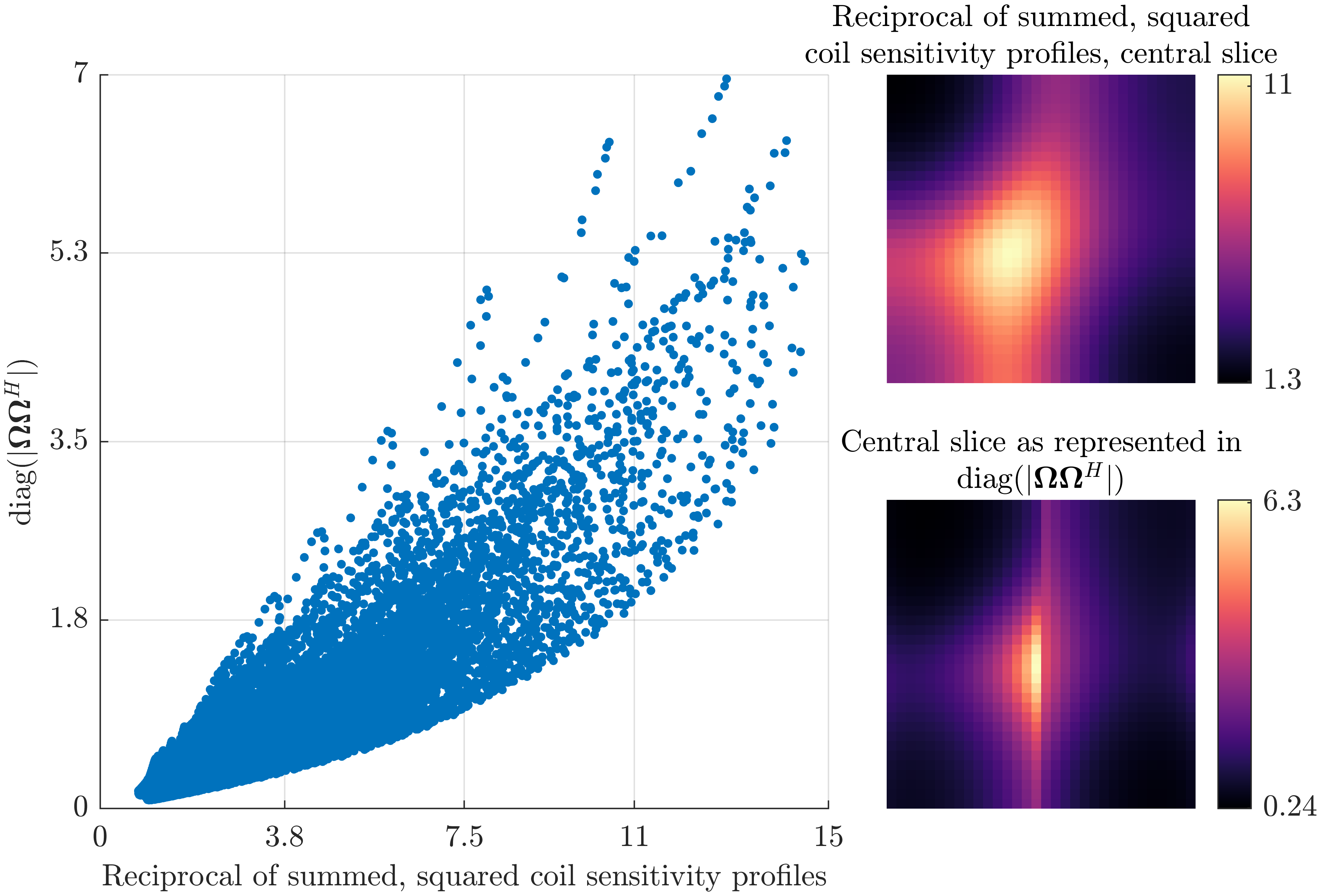}}
\caption{Analysis of summed squared coil sensitivity profiles and unnormalized temporal variance if inter image independence of the continues image time series is assumed ($\frac{1}{V}\Delta \tilde{\mymat{P}} (\Delta \tilde{\mymat{P}} )^H = \mymat{I}_{MNL}$) and scaling by $1/V$ is avoided. The scatter plot depicts these quantities for the full FOV. The two images on the side show both quantities individually restricted to the central slice of the FOV. For this example, $\mymat{R}$ represents a SENSE reconstruction (undersampling factor $R=2$ in primary phase encoding direction) such that an $L=32\times32\times 16$ image could be reconstructed. K-space was segmented in $N=16$ planes, each containing $M=32\times 16$ samples, simulating an undersampled 3D EPI. Coil sensitivity profiles of $\Lambda=8$ coils were considered.}
\label{fig_totSENSE_vs_var}
\end{figure}


\newpage
\section{Complementary properties of the spectral and total discrete timing operators}
The spectral timing operators $\hat{\mymat{H}}(f)$, $\hat{\mymat{H}}^{(Opt)}$, $\hat{\mymat{H}}^{(\varepsilon)}$, and the total discrete timing operators $\mymat{\Omega}$, $\mymat{\Omega}^{(Opt)}$, $\mymat{\Omega}^{(\varepsilon)}$, are two sets of three operators used to describe the actual, optimal, and error image time series in the context of the timing problem in 3D fMRI. Both sets of operators depend on the reconstruction matrix $\mymat{R}$ and its pseudo inverse $\mymat{R}^\dagger$. \\
The spectral operators act in the continuous spectral domain. Using $\hat{\mymat{H}}^{(\varepsilon)}$, a simple proxy for the strength of the timing error can be defined (see \myref{Section }{sec:timing_proxy_strength_error}). This proxy depends only on the sequence timing, is independent of $\mymat{R}$, and has a closed-form expression (see \myref{Equation }{eq:proxy_timingproblem_final}). It provides a crude but efficient way to assess the impact of the timing problem. \\
In contrast, the total discrete timing operators offer a more detailed characterization. However, they are very large matrices and computing, for example, their singular value decomposition can be computationally demanding. These operators act in discrete time and allow the definition and analysis of the spatial, temporal, and total variance of the image time series.
\newpage
\section{Simulations}\label{sec_simulations}
In the previous section, a formal model was derived for the timing problem of 3D fMRI data.
In this section, two simulations are conducted using the derived formalism. We aim to foster
a deeper and more practical understanding of the timing problem using these simulations. In
\myref{Section }{subsec_osc_ps_sim}, a point source is simulated that oscillates with increasing frequency. Describing
the depiction of point sources is a standard procedure when studying imaging systems. The
simulation presented in \myref{Section }{subsec_bl_object} closely based on the experiments presented by \cite{vanderZwaag2012}. Using this simulation, we can study what the timing problem implies for the
stability of 3D fMRI image time series. As shown in \myref{Equation }{eq_vtot_svd_based}, the total variance ($\nu^{(Tot)}$)
depends on the singular values of the total discrete timing operator and the representation of
the continues image time series in the column space of the total discrete timing operator. The results of this interaction will be investigated in the following.

\subsection{Imaging an oscillating point-source}\label{subsec_osc_ps_sim}
We simulate an oscillating sinusoidal point source in the center of the FOV (indexed by $l_c$). Therefore, the continues image is generated according to:
\begin{equation}\label{eq:timing_sinusoidal_point_source}  
    \Delta\myvec{\varrho}^{(c)}_{l}(\tau_{(m,n,0)}) = 
    \begin{cases}
        e^{2 \pi i f \tau_{(m,n,0)}} & \text{if } l=l_c \\
        0 & \text{else}
    \end{cases}
\end{equation}
Applying $\mymat{\Omega}$ results in the image $\Delta \mymat{\varrho}$, applying $\mymat{\Omega}^{(\varepsilon)}$ results in the error image $\Delta \mymat{\varrho}^{(\varepsilon)}$, and applying $\mymat{\Omega}^{(Opt)}$ results in the optimal image $\Delta \mymat{\varrho}^{(Opt)}$. The acquisition of one image was simulated for different frequencies $f$. $f$ was ramped from $0$ to $8/{2 T_{Vol}}$. Maximum-intensity projections (MIPs) of the imaged point source are presented in \myref{Figure }{fig_mip_oscilating_point_source}.  
\\In the DFT-based simulation, the reconstructed signal intensity shifts away from the point source's original, central location as the oscillation frequency $f$ increases. This behavior is fully in line with what could be predicted using the sensitivity fields presented in \myref{Figure }{fig_timing_sense_field_ex} and underlines that the timing operatros act as spatio-temporal filters. The simulated frequency range ($f \in [0, 8/2T_{Vol}]$) is too small to cause appreciable timing artifacts along the measurement (M) or primary phase encoding (P) directions. Instead, the artifacts occur almost entirely along the secondary phase encoding (S) direction, which corresponds to the segmentation of k-space. The timing error image $\Delta \varrho^{(\varepsilon)}$ demonstrates sensitivity to both: false signal intensity appearing at incorrect spatial locations and the corresponding signal missing intensity at the original location of the point-source.
\\In the SENSE-based simulation, the reconstructed signal intensity follows very similar shifting pattern as the oscillation frequency $f$ increases, but its spatial distribution is modulated due to the inclusion of coil sensitivity profiles in the reconstruction. The strength of the timing error slightly decreases towards the edge of the FOV due to the interplay of sequence timing and parallel imaging in the simulated scenario. Subtle timing artifacts appear also along the primary phase encoding direction (P) (around index $-16$).
\\In summary, both simulations motivate the conclusion that each voxel possesses a unique, spatio-temporal point-spread function governed by the interaction between the timing operator and the temporal signal dynamics during acquisition.
\begin{figure}
\centerline{\includegraphics[width=17cm]{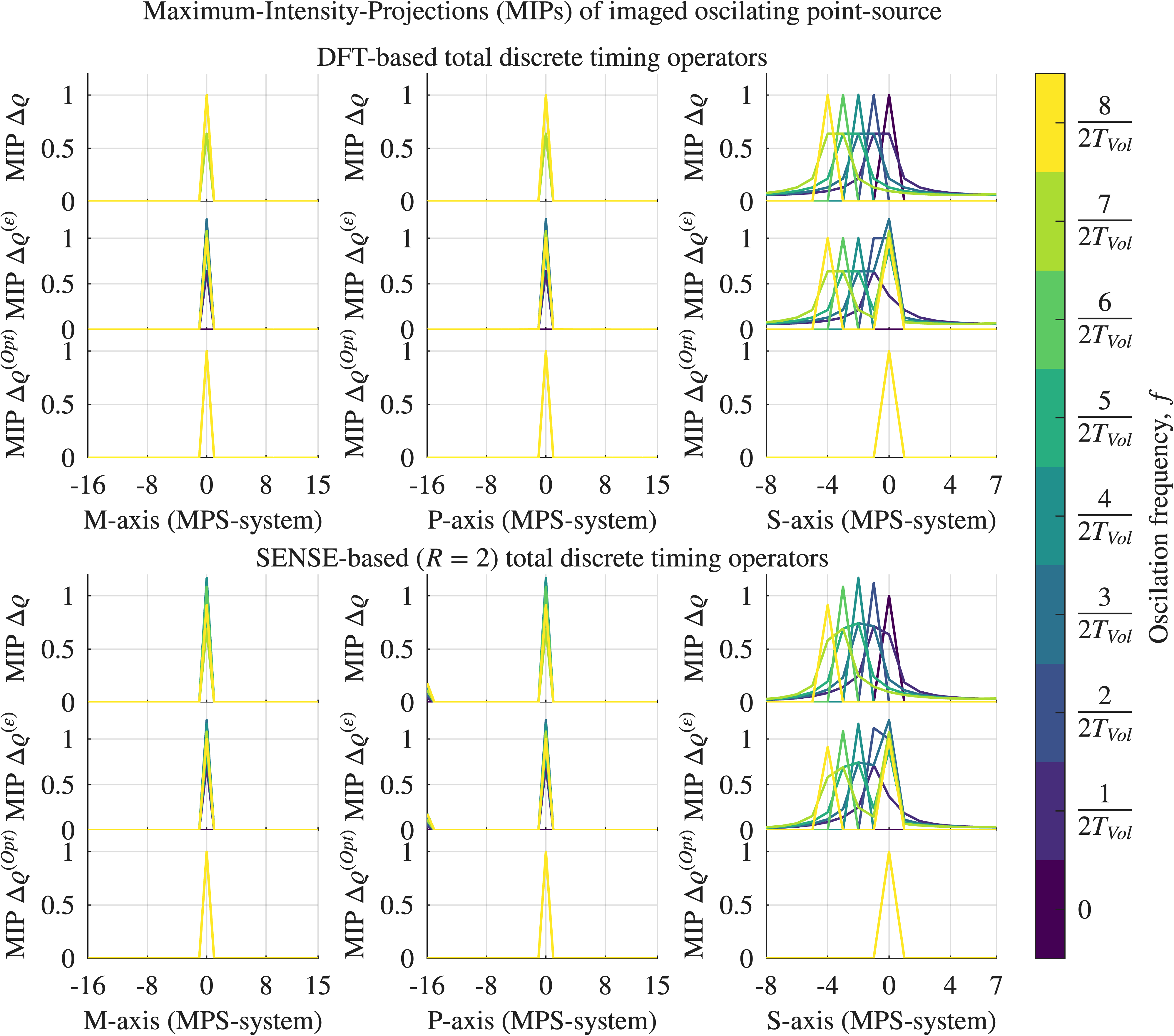}}
\caption{Maximum-intensity projections (MIPs) of a simulated, oscillating point source in the center of the FOV. $f$ specifies the oscilation frequency of the point source. For the upper plots, all discrete timing operators ($\mymat{\Omega}$,$\mymat{\Omega}^{(\varepsilon)}$,$\mymat{\Omega}^{(Opt)}$) were based on $\mymat{R}$ representing a inverse 3D DFT such that an $L=32\times32\times 16$ image could be reconstructed. K-space was segmented in $N=16$ planes, each containing $M=32\times 32$ samples, simulating a fully sampled 3D EPI. No coil sensitivity profiles were considered ($\Lambda = 1$). $T_R$ was set to $60ms$, $T_S$ to $20ms$, $T_D$ to $20ms/2^{14}$, and $T_{Vol}=NT_R = 0.96s$. For the lower plots, all discrete timing operators were based on $\mymat{R}$ representing  a SENSE reconstruction (undersampling factor $R = 2$ in primary phase encoding direction) such that an $L = 32 \times 32 \times 16$ image could be reconstructed. K-space was segmented in $N = 16$ planes, each containing $M = 32 \times 16$ samples, simulating an undersampled 3D EPI. Coil sensitivity profiles of $\Lambda = 8$ were considered. $T_R$ was set to $60ms$, $T_S$ to $20ms$, $T_D$ to $20ms/2^{14}$, and $T_{Vol}=NT_R = 0.96s$.}
\label{fig_mip_oscilating_point_source}
\end{figure}

\subsection{Imaging a band-limited object}\label{subsec_bl_object}

For all results presented in this section, we simulated a continues image $\Delta \myvec{\varrho}^{(c)}(\tau)$ with voxel times series generated by sampling band-limited white noise. We chose $8$ band limits equally distributed in the interval $[1/(2\cdot 16T_R),8/(2 \cdot 16T_R)]$ and $1$ band limit equal to $\infty$. The image always has dimensions $32\times32\times N$ and the acquired k-space consists of $N$ planes/segments containing $M=32\times 32$ (DFT-case) or $M=32\times 16$ (SENSE-case) samples, effectively simulating 3D EPI with $N$ RF cycles necessary per image. $N$ was ramped from $1$ to $16$. The reconstruction matrix $\mymat{R}$ was based on a 3D DFT or a SENSE reconstruction including $\Lambda = 8$ coil sensitivity profiles and with undersampling factor $2$ in primary phase encoding direction. $T_R$ was always set to $60ms$, $T_S$ to $20ms$ and $T_D$ to $20ms/2^{14}$. For each simulation $V=100$ images were simulated. The image time series $\Delta\myvec{\varrho}$, $\Delta \myvec{\varrho}^{(Opt)}$, and $\Delta \myvec{\varrho}^{(\varepsilon)}$ could easily be computed by applying $\mymat{\Omega}$, $\mymat{\Omega}^{(Opt)}$ or $\mymat{\Omega}^{(\varepsilon)}$, respectively, to the generated $\Delta \tilde{\mymat{P}}$. This simulation, especially when using the SENSE reconstruction, closely resembles the experiments described by \cite{vanderZwaag2012}. However, we had to restrict the image dimensions to make the simulation computationally feasible. All simulated image times series can be analyzed and compared in terms of their total variance. To assess the reproducibility of our results, $25$ different continues images per band limit were generated. All results are summarized in \myref{Figure }{fig_bandlimited_noise_sim}.
\\For the DFT-based reconstruction, the normalized total variance of the reconstructed image time series ($\nu^{(\mathrm{Tot})}(\mymat{\Delta \tilde{P}}; \mymat{\Omega}) / \nu^{(\mathrm{Tot})}(\mymat{\Delta \tilde{P}}; \mymat{\Omega}^{(\mathrm{Opt})})$) remains close to $1$, virtually independent of both the number of segments $N$ and the noise frequency band limit $f$. In contrast, the variance ratio of the timing error image time series ($\Delta \varrho^{(\varepsilon)}$) increases with $N$ and $f$, reaching a value close to $2$.  In the non-band-limited noise case ($f = \infty$) the mean curve remains approximately equal to $2$. Notably, the shape of the error variance ratio curves closely resembles the frequency and  number of segments dependence predicted by the proxy for the strength of the timing error, $\epsilon(f)$ (see \myref{Section }{sec:timing_proxy_strength_error}).
\\In the SENSE-based reconstruction, the total variance ratio for the image time series decreases as $N$ and $f$ increase, yielding the lowest mean curve when the band limit is infinite ($f = \infty$). The timing error variance ratio increases with higher $f$ and $N$, though this effect is substantially less pronounced compared to the DFT case. Furthermore, the error variance ratio exhibits a slight decrease for the non-band-limited case ($f = \infty$). 
\\The total variance ratios serve as a global measure of image time series stability relative to the timing optimal image time series ($\Delta \varrho^{(\mathrm{Opt})}$). Across both reconstruction types, the obtained relative error variance curves align closely with the timing error proxy $\epsilon(f)$. In the SENSE case, the observed increase in time series stability (lower variance ratio) with higher $N$ and $f$ is consistent with the decreasing singular values of $\mymat{\Omega}$ as $N$ grows. Overall, the error image time series exhibits greater instability than the reconstructed image time series, which aligns directly with the larger singular values of the timing error operator $\mymat{\Omega}^{(\varepsilon)}$. 
\begin{figure}
\centerline{\includegraphics[width=17cm]{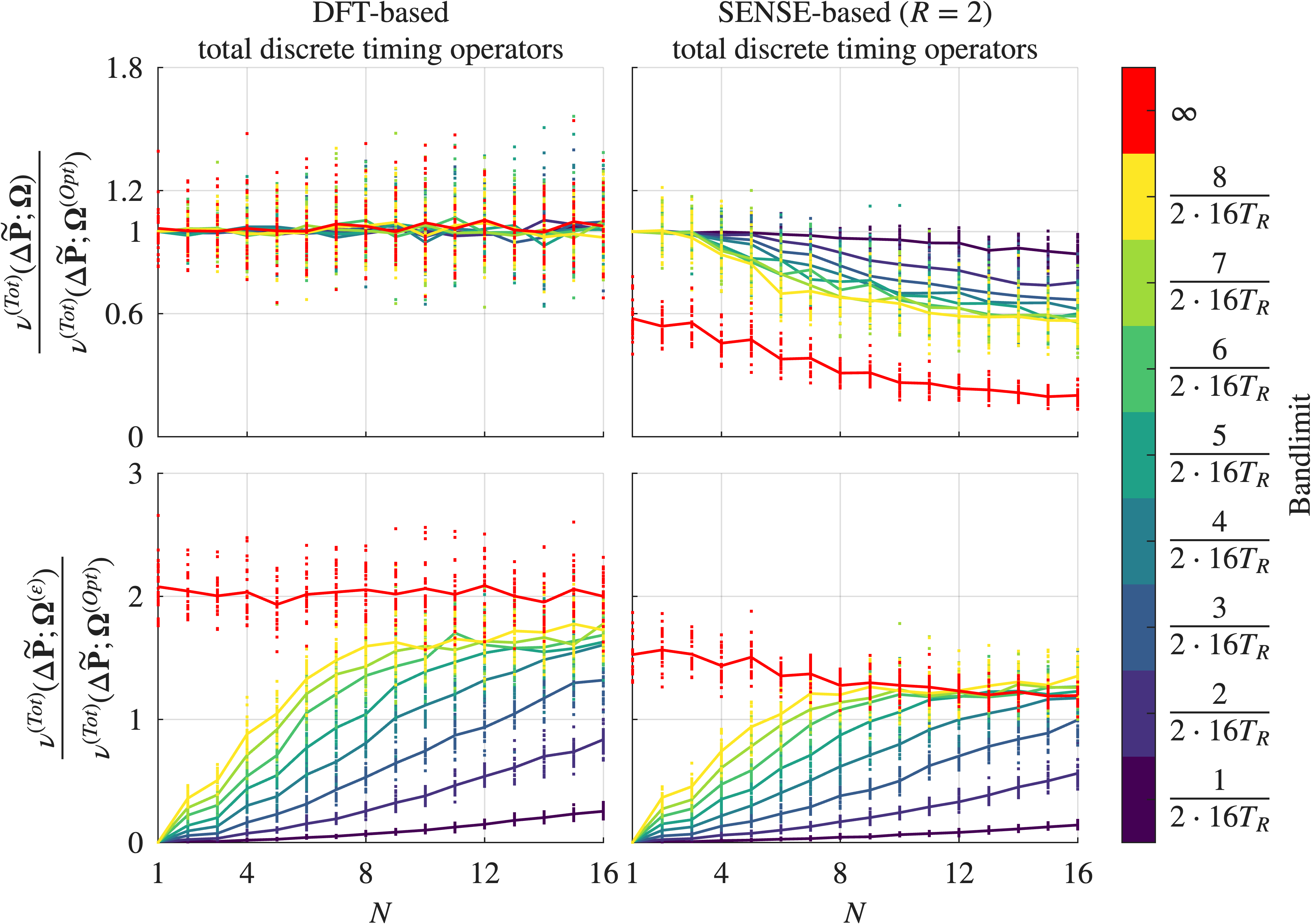}}
\caption{Total variance of simulated image time series $\nu^{(Tot)}(\Delta \tilde{\mymat{P}}; \mymat{\Omega})$ and error image timeseries $\nu^{(Tot)}(\Delta \tilde{\mymat{P}}; \mymat{\Omega}^{(\varepsilon)})$ normalized by the total variance of the optima image time series $\nu^{(Tot)}(\Delta \tilde{\mymat{P}}; \mymat{\Omega}^{(Opt)})$. All simulated images, within a timeseries, had dimensions of $32 \times 32 \times N$, modeling a 3D EPI acquisition requiring $N$ RF excitation cycles per volume (with $N$ ramped from $1$ to $16$). The acquired k-space comprised $N$ segments containing $M = 32 \times 32$ samples for the DFT-case and $M = 32 \times 16$ samples for the SENSE-case. The reconstruction matrix $\mymat{R}$ represented either a inverse 3D DFT or a SENSE reconstruction including $\Lambda = 8$ coil sensitivity profiles and an undersampling factor of $2$ in the primary phase-encoding direction. $T_R=60ms$, $T_S=20ms$, and $T_D=20ms/2^{14}$ were used and each simulation run generated a time series of $V = 100$ images. $25$ independent continuous image realizations were generated for each run by sampling band-limited white noise (bandlimits are indicated on the colorbar). Dots correspond to results of individual simulation runs and lines represent the mean over all $25$ runs.}
\label{fig_bandlimited_noise_sim}
\end{figure}
\newpage
\section{Time series analysis using GLMs}\label{sec_timing_ts_analysis_GLM}
The model presented for the timing of 3D fMRI data has some implications for the analysis of image time series using GLMs. Typically, GLM analyses are performed on the magnitude of the image time series. However, to fully exploit the linearity of the total discrete timing operators, all considerations presented here are based on the complex-valued images. \\
The design matrix $\mymat{X}$ is usually constructed without considering the timing problem when analyzing 3D fMRI data. Therefore, we can think of the design matrix as being created with the timing-optimal image time series, $\Delta \mymat{\varrho}^{(Opt)}$, in mind. Using the GLM framework this leads to: 
\begin{equation}
    \Delta \mymat{\varrho}^{(Opt)} = \mymat{B}^{(Opt)}\mymat{X} + \mymat{\eta}^{(Opt)}
\end{equation}
Note that, $\Delta \mymat{\varrho}^{(Opt)}$ is an $L\times V$ matrix, and the voxel index $l$ refers to its rows and the image index $v$ refers to its columns. Therefore, the notation used here must deviate from the typical notation used for GLMs, $\myvec{y} = \mymat{X}\myvec{\beta} + \myvec{\eta}$ (see \cite{Friston1994}), where the image index refers to rows of $\myvec{y}$ and each voxel is considered individually. $\mymat{\eta}^{(Opt)}$ is the error term. Typically, it is assumed that $\mymat{\eta}^{(Opt)}$ contains independent and identically distributed noise samples. \\
$\Delta \tilde{\mymat{P}}$, the matrix containing the continues image time series (see \myref{Equation }{eq:timing_full_continues_image}), can also be modeled using the same design matrix $\mymat{X}$. 
\begin{equation}\label{eq_glm_cont_image_1}
    \Delta \tilde{\mymat{P}} = \tilde{\mymat{B}} \mymat{X} + \tilde{\mymat{\eta}}
\end{equation}
In this equation, $\mymat{X}$ remains a $P\times V$ matrix, where $P$ equals the number of regressors. Consequently, $\tilde{\mymat{B}}$ is a $MNL\times P$ matrix. It contains one GLM parameter per voxel, regressor, and, additionally, intra image time point. Using the total discrete timing operators, it can be shown directly that the timing problem will certainly lead to errors in the parameters $\mymat{B}$.
\begin{equation}
    \mymat{B} = \mymat{\Omega}\tilde{\mymat{B}} \quad 
    \mymat{B}^{(Opt)} = \mymat{\Omega}^{(Opt)}\tilde{\mymat{B}}  \quad
    \mymat{B} - \mymat{B}^{(Opt)} = \mymat{B}^{(\varepsilon)} = \mymat{\Omega}^{(\varepsilon)}\tilde{\mymat{B}} 
\end{equation}
Based on everything that has been discussed so far in this chapter, it is immediately clear that the error $\mymat{B}^{(\varepsilon)}$ increases in magnitude with the band limit of the intra image fluctuations present in $\Delta \tilde{\mymat{P}}$. Consequently, and even if $\mymat{X}$ fits $\Delta \tilde{\mymat{P}}$ as well as $\Delta \mymat{\varrho}^{(Opt)}$, the GLM parameters $\mymat{B}$ may still be distorted by the timing problem and differ from the optimal parameters $\mymat{B}^{(Opt)}$. \\
In general, we are unable to make a definitive statement regarding the impact of the timing problem on the quality of the GLM fit. However, assuming that the simulation results displayed in \myref{Figure }{fig_bandlimited_noise_sim} generalize sufficiently well, we can outline a scenario in which the timing problem might actually improve the quality of the fit. Although this may seem counterintuitive, since the timing problem inevitably introduces errors into the estimated GLM parameters, our reasoning is based on two observations from the simulation results:
\begin{itemize}
    \item The ratio $\nu^{(Tot)}(\Delta \tilde{\mymat{P}};\mymat{\Omega})/\nu^{(Tot)}(\Delta \tilde{\mymat{P}};\mymat{\Omega}^{(Opt)})$ is smaller or equal to $1$
    \item In the SENSE case (\myref{Figure }{fig_bandlimited_noise_sim}), this ratio decreases with increasing $f_{Max}$. Hence, we expect it to be minimal in the case of pure white noise ($f_{Max} = \infty$). 
\end{itemize}
The quality of the GLM fit can be measured using the $R^2$. In the timing-optimal case we find:
\begin{equation}
    R^{2(Opt)} = \frac{\nu^{(Tot)}(\tilde{\mymat{B}} \mymat{X};\mymat{\Omega}^{(Opt)})}
    {\nu^{(Tot)}(\tilde{\mymat{B}} \mymat{X};\mymat{\Omega}^{(Opt)}) + 
    \nu^{(Tot)}(\tilde{\mymat{\eta}};\mymat{\Omega}^{(Opt)})}
\end{equation}
Here, it was assumed that $\tilde{\mymat{B}}$ coincides with its ordinary least squares estimate. Similarly, we can write: 
\begin{equation}
    R^{2} = \frac{\nu^{(Tot)}(\tilde{\mymat{B}} \mymat{X};\mymat{\Omega})}
    {\nu^{(Tot)}(\tilde{\mymat{B}} \mymat{X};\mymat{\Omega}) + 
    \nu^{(Tot)}(\tilde{\mymat{\eta}};\mymat{\Omega})}
\end{equation}
If, relative to the timing-optimal case, the total variance of $\tilde{\mymat{\eta}}$ is attenuated more strongly by $\mymat{\Omega}$ than the total variance of $\tilde{\mymat{B}} \mymat{X}$, then the resulting $R^2$ will be greater than $R^{2(Opt)}$. Formally, this scenario can be expressed as:
\begin{equation}
    1 \geq \frac{\nu^{(Tot)}(\tilde{\mymat{B}} \mymat{X};\mymat{\Omega})}{\nu^{(Tot)}(\tilde{\mymat{B}} \mymat{X};\mymat{\Omega}^{(Opt)})} > \frac{\nu^{(Tot)}(\tilde{\mymat{\eta}};\mymat{\Omega})}{\nu^{(Tot)}(\tilde{\mymat{\eta}};\mymat{\Omega}^{(Opt)})} \Longrightarrow R^{2} > R^{2(Opt)}
\end{equation}
Moreover, the more $\tilde{\mymat{\eta}}$ consists of pure white noise samples without residual autocorrelation, the more its total variance will be damped by $\mymat{\Omega}$. Consequently, the better $\mymat{X}$ models $\Delta\tilde{\mymat{P}}$, the lower the residual autocorrelation of $\tilde{\mymat{\eta}}$, and the greater the potential inflation of $R^2$ due to the timing problem. Finally, this inflation of the $R^2$ may increase with $N$, the number of k-space segments. As shown in \myref{Figure }{fig_bandlimited_noise_sim}, the relative damping of the total variance becomes more pronounced as $N$ increases.
\newpage
\section{Discussion}\label{sec_discussion}
In \myref{Sections }{sec_isolating_timing_problem}, \myref{}{sec_spectral_timing_operators}, and \myref{}{sec_total_discrete_timing_operators} we proposed a model for the timing of 3D fMRI data, and in \myref{Section }{sec_simulations} we studied the properties of this model based on simulated data. We also analyzed the implications of our model for GLM-based time series analysis in \myref{Section }{sec_timing_ts_analysis_GLM}. Here, we want to bring the discussion back the five questions we formulated at the beginning of this chapter (see \myref{Section }{sec_intro}). Using the derived formalism, we attempt to answer them as clearly as possible. \\
\textbf{\textit{Can 3D fMRI data be used to infer on effects which exceed the band limit given by the Nyquist-Shannon sampling theorem?}} \\
Our results clearly demonstrate that the error induced by the timing problem increases with the band limit of the fluctuations in the object, up to a saturation point. This effect is predicted by proxy for the strength of the error (see \myref{Section }{sec:timing_proxy_strength_error}) and is consistently observed in all simulations presented in \myref{Section }{sec_simulations}. As discussed in \myref{Section }{sec_timing_ts_analysis_GLM}, this error inevitably affects the parameter estimates of any GLM. Consequently, accurately inferring signal components with a band limit exceeding $1/2T_{Vol}$ is inherently error prone. The simulated PSFs in \myref{Figures }{fig_mip_oscilating_point_source} illustrate that the timing problem distorts the time series within the voxel of origin and introduces additional signal content into neighboring voxels. As a result, GLM parameter estimates can be altered both in magnitude and spatial distribution. \\
\textbf{\textit{Can the decline of tSNR with increasing number of k-space segments be exclusively attributed to physiological processes?}} \\ 
The decline of tSNR with an increasing number of k-space segments has been very clearly demonstrated by \cite{vanderZwaag2012}. Thus, we expected to easily reproduce this effect in our simulations, even though we simulate band-limited white noise rather than, for example, realistic physiological noise. Interestingly, our findings, based on simulations of the timing of 3D fMRI data, appear to contradict the observations reported by \cite{vanderZwaag2012}. If the timing problem introduced additional instabilities into the image time series, the ratio of $\nu^{(Tot)}(\Delta \tilde{\mymat{P}};\mymat{\Omega})$ to $\nu^{(Tot)}(\Delta \tilde{\mymat{P}};\mymat{\Omega}^{(Opt)})$ (i.e., the total variance of the actual image time series relative to that of the timing-optimal image time series) should have increased, particularly when k-space is divided into many segments. However, our simulations show that $\nu^{(Tot)}(\Delta\tilde{\mymat{P}};\mymat{\Omega})/\nu^{(Tot)}(\Delta \tilde{\mymat{P}};\mymat{\Omega}^{(Opt)})$ 
remains unchanged or even decreases as the number of k-space segments increases (see \myref{Figure }{fig_bandlimited_noise_sim}). This suggests that the global stability of the image time series may actually improve as a result of the timing problem. Assuming that these results generalize to any physiological process and imaging technique, and under the initial assumptions outlined in \myref{Section }{sec_isolating_timing_problem}, we must conclude that the decline of tSNR cannot be attributed to physiological processes. As a reminder, these assumptions state that the parameters governing transverse magnetization dynamics after RF excitation change over time and that linear image reconstruction is performed. \\
Subject motion and scanner instabilities offer alternative explanations for the observed tSNR decline. Early findings, presented in \cite{Reidel2026}, suggest that accounting for both motion and scanner instabilities can improve tSNR. Intuitively, it is expected that the tSNR improvements reported by \cite{Riedel2026} will become even more pronounced as k-space is divided into more segments. \\
\textbf{\textit{Why is 3D fMRI so sensitive to physiological noise?}} \\
We can reconcile this observation with our framework, although the explanation deviates from the prevailing interpretations in the literature. In our simulations, we demonstrate that the total variance of band-limited white noise is damped by the total discrete timing operator, with stronger damping observed at higher band limits. If we assume that realistic physiological noise behaves similarly to band-limited white noise, then its total variance will also be reduced by the timing problem. However, white noise with full bandwidth, such as thermal noise, is damped even more strongly. As a result, the relative contribution of physiological noise to the total variance of the image time series actually increases. This effect becomes more pronounced as the number of k-space segments increases. The more accurately physiological noise is modeled in a GLM analysis, the larger the fraction of the unexplained variance that truly originates from purely whit noise sources, which is damped the most by the total discrete timing operator. In contrast to the common view in the existing literature, where it is suspected that k-space segmentation "amplifies" physiological noise, our simulations do not support this interpretation.
\textbf{\textit{Why plays RETROICOR such a crucial role in the physiological noise correction of 3D fMRI data?}} \\
As discussed in \myref{Section }{sec_timing_ts_analysis_GLM}, the better the design matrix of a GLM models temporal intra image fluctuations, the greater the potential inflation of the $R^2$. RETROICOR models physiological noise by employing a Fourier expansion of the cardiac and respiratory phases. The use of the Fourier expansion results in the generation of regressors which effectively capture the associated intra-image fluctuations. Therefore, RETROICOR performs exceptionally well in modeling physiological noise in 3D fMRI data, although the timing problem alters the spatiotemporal pattern of physiological noise in the image time series. This explanation is further supported by the fact that RETROKCOR is as efficient in correcting for physiological noise in the data as RETROICOR, as shown in \cite{Tijssen2014}. \\
\textbf{\textit{Why should regressors be sampled at time points coinciding with the acquisition of the temporally central k-space segment?}} \\ 
This finding aligns with the optimal choice of $T'$ derived using the proxy for the strength of the timing \myref{Equation }{eq:porxy_timingporblem}). Notably, the proxy predicts this optimal $T'$ regardless of the specific k-space trajectory. Thus, even if the center of k-space is sampled during the, for example, first RF cycle, $T'$ should still be selected such that it lies at the center of the acquisition window for a single image. 
\subsubsection{Final remarks}
We successfully derived a model for the timing of 3D fMRI data. Importantly, we can fuse the model with many observations that are documented in the literature. Hence, the model is a relevant contribution to the field. However, aspects such as the sensitivity fields (see \myref{Equation }{eq:timing_def_sf}) or
the model's implications for the covariance structure of the fMRI data (see "Temporal variance" in \myref{Section }{sec:timing_mapping_props_of_omega}) have been defined but not explored in the context of the existing literature. We hope that the general discussion of the model and its properties helps to foster a general understanding of the timing problem in 3D fMRI that might be relevant for future research in the field.

\section{Appendix}

\subsection{Derivation of the proxy in \myref{Equation }{eq:porxy_timingporblem}}\label{timing_proofs_3}
We start with
\begin{equation}
\begin{gathered}
    \left|e^{2 \pi i f \tau_{(m,n,0)}}-e^{2 \pi i f T'} \right|^2 = 
    \left|e^{2 \pi i f (\tau_{(m,n,0)}-T')}-1 \right|^2 = \\
    \left(e^{2 \pi i f (\tau_{(m,n,0)}-T')}-1 \right)\left(e^{-2 \pi i f (\tau_{(m,n,0)}-T')}-1 \right) = \\
    1-e^{2 \pi i f (\tau_{(m,n,0)}-T')} - e^{-2 \pi i f (\tau_{(m,n,0)}-T')}+1 = \\
    2 - 2\frac{e^{2 \pi i f (\tau_{(m,n,0)}-T')} + e^{-2 \pi i f (\tau_{(m,n,0)}-T')}}{2} = 
    2-2\cos(2 \pi  f (\tau_{(m,n,0)}-T'))
\end{gathered}
\end{equation}
The sum can be rewritten as
\begin{equation}
    \sum_{(m,n)}{\left|e^{2 \pi i f \tau_{(m,n,0)}}-e^{2 \pi i f T'} \right|^2} = 2MN - 2 
    \sum_{(m,n)}{\cos(2 \pi  f (\tau_{(m,n,0)}-T'))}
\end{equation}
The cosine can be formulated as
\begin{equation}
    \cos(2 \pi f(\underbrace{T_S + m T_D +n T_R}_{=\tau_{(m,n,0)}} - T')) = \text{Re}\left( e^{2\pi i f(T_s-T')}e^{2\pi i fnT_R}e^{2\pi i fmT_D}\right)
\end{equation}
Hence, the sum over the cosine is given as
\begin{equation}
    \sum_{(m,n)}{\cos(2 \pi  f (\tau_{(m,n,0)}-T'))} = \text{Re}\left( e^{2\pi i f(T_s-T')}\sum_n{\left(e^{2\pi i fnT_R} \sum_{m}{e^{2\pi i fmT_D}}\right)}\right)
\end{equation}
In the following, we will use the closed form of the geometric sum. 
\begin{equation}
\begin{gathered} 
    \sum_{m}{e^{2\pi i fmT_D}} = \sum_{m}{\left(e^{2\pi i fT_D}\right)^m} =
    \frac{1-e^{2\pi i fMT_D}}{1-e^{2\pi i fT_D}} =  \\
    \frac{e^{\pi i f MT_D}\left(e^{-\pi i f MT_D} - e^{\pi i f MT_D} \right)}{e^{\pi i fT_D}
    \left(e^{-\pi i f T_D} - e^{\pi i f T_D} \right)} = 
    \frac{e^{\pi i f MT_D}}{e^{\pi i f T_D}} \frac{\sin(\pi f M T_D)}{\sin(\pi f T_D)} = \\
    e^{\pi i f (M-1)T_D}\frac{\sin(\pi f M T_D)}{\sin(\pi f T_D)}
\end{gathered}
\end{equation}
As this results can be moved in front of the sum over $n$. The sum over $n$ can be written as
\begin{equation}
\begin{gathered} 
    \sum_{n}{e^{2\pi i fnT_R}} = 
    e^{\pi i f (N-1)T_R} \frac{\sin(\pi f N T_R)}{\sin(\pi f T_R)}
\end{gathered}
\end{equation}
Bringing everything back together, we find
\begin{equation}
\begin{gathered} 
    \text{Re}\left( e^{2\pi i f(T_s-T')}\sum_n{\left(e^{2\pi i fnT_R} \sum_{m}{e^{2\pi i fmT_D}}\right)}\right) = \\
    \text{Re}\left(e^{\pi i f (2(T_S-T')+(M-1)T_D + (N-1)T_R)}\frac{\sin(\pi f N T_R)}{\sin(\pi f T_R)}\frac{\sin(\pi f M T_D)}{\sin(\pi f T_D)}\right) = \\
    \cos(2\pi f(T_S-T') + \pi f (M-1)T_D + \pi f (N-1)T_R) \frac{\sin(\pi f N T_R)}{\sin(\pi f T_R)}\frac{\sin(\pi f M T_D)}{\sin(\pi f T_D)}
\end{gathered}
\end{equation}
Using the definition of the Chebyshev polynomials of the second kind we obtain
\begin{equation}
    U_{M-1}\left(\cos(\pi f T_D)\right)U_{N-1}\left(\cos(\pi f T_R)\right)  \cos(2\pi f(T_S-T') + \pi f (M-1)T_D + \pi f (N-1)T_R)
\end{equation}
Subtracting this from $2MN$ and rearranging the terms leads to \myref{Equation }{eq:porxy_timingporblem}.

\subsection{Time-domain form of \myref{Equation }{eq:timing_operator_applied}}\label{timing_proofs_1}
We can use the following two equivalent formulations
\begin{equation}
    \hat{\mymat{H}}(f)\Delta \hat{\varrho}^{(c)}(f) \Longleftrightarrow \sum_{l'}{\hat{\mymat{H}}_{l,l'}(f)\Delta \hat{\varrho}_{l'}^{(c)}(f) }
\end{equation}
\begin{equation}
\hat{\mymat{H}}(f) = \mymat{R} \left(\mymat{I}_{\Lambda}\otimes \hat{\mymat{D}}(f)\right) \mymat{R}^{\dagger} \Longleftrightarrow \hat{\mymat{H}}_{l,l'}(f) = \sum_{(m,n,\lambda)}{\mymat{R}_{l,(m,n,\lambda}e^{2 \pi i f \tau_{(m,n,0)}}\mymat{R}^\dagger_{(m,n,\lambda),l'}}
\end{equation}
\myref{Equation }{eq:timing_operator_applied} can, therefore be written as
\begin{equation}
    \Delta\hat{\varrho}_l(f) = \left(\sum_{l'}{\sum_{(m,n,\lambda)}{ \mymat{R}_{l,(m,n,\lambda)}e^{2 \pi i f \tau_{(m,n,0)}}\mymat{R}^\dagger_{(m,n,\lambda),l'} \Delta \hat{\varrho}_{l'}^{(c)}(f) }}\right) * \sum_{v=-\infty}^{\infty}{\delta\left(f-v/T_{Vol}\right)}
\end{equation}
Applying the inverse Fourier transform leads to: 
\begin{equation}
    \Delta \varrho_{l,v} = \sum_{l'}{\sum_{(m,n,\lambda)}{\mymat{R}_{l,(m,n,\lambda)} \Delta\varrho_{l'}^{(c)}(\tau_{(m,n,0)}+vT_{Vol})\mymat{R}^{\dagger}_{(m,n,\lambda),l'}}}
\end{equation}
For this step, we used that the inverse Fourier transform of
\begin{equation}
    \left(e^{2 \pi i f \tau_{(m,n,0)}}\Delta \hat{\varrho}_{l'}^{(c)}(f)\right) * \sum_{v=-\infty}^{\infty}{\delta\left(f-v/T_{Vol}\right)}
\end{equation}
equals $\Delta\varrho_{l'}^{(c)}(\tau_{(m,n,0)}+vT_{Vol})$. 

\subsubsection{Proof of \myref{Equation }{eq:timing_total_discrete_op_def}}\label{timing_proofs_2}
All we need to show is 
\begin{equation}
    \left(\mymat{R}^{\dagger} \circ \left(\myvec{1}_\Lambda \otimes (\Delta\mymat{P}^{(v)})^T \right)\right) \myvec{1}_L = 
    \left(\left( \myvec{1}_\Lambda  \otimes \mymat{I}_{MN}\right) \bullet \mymat{R}^\dagger\right) \text{vec}\left(\Delta\mymat{P}^{(v)}\right)
\end{equation}
The equality is preserved when left-multiplying $\mymat{R}$. For the proof we need
\begin{equation}
    \text{diag}\left(\mymat{A}\right) = \left( \mymat{I} \bullet \mymat{I} \right)\text{vec}\left(\mymat{A}\right)
\end{equation}
\begin{equation}
    \left(\mymat{A} \circ \mymat{B}^T\right)\myvec{1} = \text{diag}\left(
    \mymat{A}\mymat{B}\right)
\end{equation}
\begin{equation}
    \text{vec}\left(\mymat{A}\mymat{C}\mymat{B}\right) = \left( \mymat{B}^T \otimes \mymat{A} \right)\text{vec}\left(\mymat{C}\right)
\end{equation}
\begin{equation}
    \left(\mymat{A} \bullet \mymat{B} \right)\left(\mymat{C}\otimes \mymat{D} \right) = \left( \mymat{A} \mymat{C}\right)\bullet \left( \mymat{B} \mymat{D}\right)
\end{equation}
Using this relationships we find
\begin{equation}
\begin{gathered}
    \left(\mymat{R}^{\dagger} \circ \left(\myvec{1}_\Lambda \otimes (\Delta\mymat{P}_v^{(v)})^T \right)\right) \myvec{1}_L 
    = \text{diag}\left(\mymat{R}^\dagger \left(\myvec{1}_\Lambda^T \otimes \Delta\mymat{P}_v^{(v)} \right) \right) 
     \\ = \left( \mymat{I}_{MN\Lambda} \bullet \mymat{I}_{MN\Lambda} \right)\text{vec}\left( \mymat{R}^\dagger \left(\myvec{1}_\Lambda^T \otimes \Delta\mymat{P}_v^{(v)}  \right)\right)
     \\ = \left( \mymat{I}_{MN\Lambda} \bullet \mymat{I}_{MN\Lambda} \right) \left(\mymat{I}_{MN\Lambda} \otimes \mymat{R}^\dagger \right)\text{vec}\left(\myvec{1}_\Lambda^T \otimes \Delta\mymat{P}_v^{(v)} \right)
     \\ = \left( \mymat{I}_{MN\Lambda} \bullet \mymat{R}^\dagger \right)\text{vec}\left( \myvec{1}_\Lambda^T \otimes \Delta\mymat{P}_v^{(v)}\right)
     \\ = \left( \mymat{I}_{MN\Lambda} \bullet \mymat{R}^\dagger \right) 
          \left(\myvec{1}_\Lambda \otimes \mymat{I}_{MNL} \right)
        \text{vec}\left( \Delta\mymat{P}_v^{(v)}\right)
    \\ = \left( \mymat{I}_{MN\Lambda} \bullet \mymat{R}^\dagger \right) 
          \left(\myvec{1}_\Lambda \otimes \mymat{I}_{MN} \otimes \mymat{I}_L \right)
        \text{vec}\left( \Delta\mymat{P}_v^{(v)}\right) 
    \\ = \left(\left(\mymat{I}_{MN\Lambda}\left(\myvec{1}_\Lambda \otimes \mymat{I}_{MN} \right) \right) \bullet \left(\mymat{R}^\dagger \mymat{I}_L \right)\right) \text{vec}\left( \Delta\mymat{P}_v^{(v)}\right)
    \\ = \left(\left(\myvec{1}_\Lambda \otimes \mymat{I}_{MN} \right) \bullet \mymat{R}^\dagger\right) \text{vec}\left( \Delta\mymat{P}_v^{(v)}\right)
\end{gathered}
\end{equation}

\bibliography{references}


\end{document}